\documentclass[twocol]{ametsocV5}

\usepackage[utf8]{inputenc}
\usepackage{dirtytalk}
\usepackage{amsmath,amsfonts,amssymb,bm}
\usepackage{mathptmx}
\usepackage{newtxtext}
\usepackage{newtxmath}
\bibpunct{(}{)}{;}{a}{}{,}

\title{Using Machine Learning to Calibrate Storm-Scale Probabilistic Guidance of Severe Weather Hazards in the Warn-on-Forecast System}

\authors{Montgomery L. Flora}

\affiliation{School of Meteorology, and Cooperative Institute for Mesoscale Meteorological Studies,  University of Oklahoma, and NOAA/OAR/National Severe Storms Laboratory, Norman, Oklahoma}
\authors{Montgomery L. Flora \correspondingauthor{Montgomery L. Flora, monte.flora@noaa.gov}} 

\extraauthor{ Corey K. Potvin$^{b,c}$, Patrick S. Skinner$^{a,b,c}$, Shawn Handler$^a$, Amy McGovern$^{c,d}$}
\extraaffil{$^a$Cooperative Institute for Mesoscale Meteorological Studies, $^b$NOAA/OAR/National Severe Storms Laboratory, $^c$School of Meteorology, University of Oklahoma, and $^d$School of Computer Science, University of Oklahoma, Norman, Oklahoma}

\abstract{ 
A primary goal of the National Oceanic and Atmospheric Administration (NOAA) Warn-on-Forecast (WoF) project is to provide rapidly updating probabilistic guidance to human forecasters for short-term (e.g., 0-3 h) severe weather forecasts. Maximizing the usefulness of probabilistic severe weather guidance from an ensemble of convection-allowing model forecasts requires calibration. In this study, we compare the skill of a simple method using updraft helicity against a series of machine learning (ML) algorithms for calibrating WoFS severe weather guidance. ML models are often used to calibrate severe weather guidance since they leverage multiple variables and discover useful patterns in complex datasets.\\
\indent Our dataset includes WoF System (WoFS) ensemble forecasts available every 5 minutes out to 150 min of lead time from the 2017-2019 NOAA Hazardous Weather Testbed Spring Forecasting Experiments (81 dates). Using a novel ensemble storm track identification method, we extracted three sets of predictors from the WoFS forecasts: intra-storm state variables, near-storm environment variables, and morphological attributes of the ensemble storm tracks. We then trained random forests, gradient-boosted trees, and logistic regression algorithms to predict which WoFS 30-min ensemble storm tracks will correspond to a tornado, severe hail, and/or severe wind report. For the simple method, we extracted the ensemble probability of 2-5 km updraft helicity (UH) exceeding a threshold (tuned per severe weather hazard) from each ensemble storm track. The three ML algorithms discriminated well for all three hazards and produced more reliable probabilities than the UH-based predictions.  Overall, the results suggest that ML-based calibrations of dynamical ensemble output can improve short term, storm-scale severe weather probabilistic guidance.
}

\begin{document}
\maketitle
%%%%%%%%%%%%%%%%%%%%%%%%%%%%%%%%%%%%%%%%%%%%%%%%%%%%%%%%%%
%INTRODUCTION
%%%%%%%%%%%%%%%%%%%%%%%%%%%%%%%%%%%%%%%%%%%%%%%%%%%%%%%%%%

\section{Introduction}\label{intro}
The National Oceanic and Atmospheric Administration (NOAA) Warn-on-Forecast program [WoF; Stensrud et al. (\citeyear{Stensrud+etal2009, Stensrud+etal2013})] is tasked with providing forecasters with reliable, probabilistic severe weather hazard guidance at very short lead times (e.g., 0-3 h).  Though operational convection-allowing models (CAMs) cannot fully resolve convective processes \citep{Bryan+etal2003} or explicitly predict severe weather hazards (e.g., tornadoes, hail $>$1 in, wind gusts $>$50 kts) CAMs with $\leq$3 km horizontal grid spacing can partially resolve important storm-scale features \citep{Potvin+Flora2015}, distinguish between severe convective modes (e.g., supercell versus mesoscale convective systems;  \citealt{Davis+etal2004,Weisman+etal2008}), and provide storm diagnostics such as updraft helicity (UH). UH is a model surrogate for supercell thunderstorms, which are prolific producers of severe weather hazards \citep{Duda+Gallus2010, Smith+etal2012}. Severe weather forecast algorithms based on UH have shown skill at both next-day (e.g., \citealt{Sobash+etal2014, Sobash+etal2016}) and $O$(1 h) lead times \citep{Snook+etal2012, Yussouf+etal2013b, Yussouf+etal2013a, Wheatley+etal2015, Yussouf+etal2015, Jones+etal2016, Skinner+etal2016, Skinner+etal2018, Jones+etal2019,Flora+etal2019, Yussouf+etal2020}. Though UH is a useful severe weather predictor, it is less correlated with severe wind events than severe hail and tornado potential and is a poor predictor of severe, non-rotating thunderstorms (which are significant producers of severe wind gusts; \citealt{Smith+etal2012, Smith+etal2013}). 

A growing alternative to using CAM severe weather surrogates are machine learning (ML) models capable of producing calibrated guidance from many input predictors (e.g., Gagne et al. \citeyear{Gagne+etal2017}; \citealt{Lagerquist+etal2017, McGovern+etal2017, Cintineo+etal2014, Cintineo+etal2018,Burke+etal2019, McGovern+etal2019_blackbox, Hill+etal2020, Lagerquist+etal2020, Cintineo+etal2020, Loken+etal2020, Sobash+etal2020, Steinkruger+etal2020}). These studies range from nowcasting lead times (e.g., $\leq$1 h; \citealt{Lagerquist+etal2017,Cintineo+etal2014,Cintineo+etal2018, Lagerquist+etal2020, Cintineo+etal2020, Steinkruger+etal2020}) which leverage available observational and numerical weather prediction (NWP) data to next-day forecasts (e.g., lead times of 24-36 h) that use state-of-the-art CAM ensemble forecasts (e.g., Gagne et al. \citeyear{Gagne+etal2017}; \citealt{Burke+etal2019,Hill+etal2020, Loken+etal2020, Sobash+etal2020}). In \citet{Lagerquist+etal2017}, ML models produced skillful probabilistic severe wind predictions for radar-observed storms.  The operational NOAA/Cooperative Institute for Meteorological Satellite Studies (CIMSS) ProbSevere model \citep{Cintineo+etal2014, Cintineo+etal2018} is a naïve Bayesian classifier that reliably predicts severe weather likelihood up to a lead time of 90 min. In a newer version, ProbSevere v2.0, the system can now produce probabilistic guidance for separate severe weather hazards \citep{Cintineo+etal2020}. Using a convolution neural network (CNN; \citealt{LeCun+etal1990}), a deep learning technique, \citet{Lagerquist+etal2020} produced a next-hour tornado prediction system with skill comparable to the ProbSevere system. In an idealized framework, \citet{Steinkruger+etal2020} explored using ML methods to produce automated tornado warning guidance and found promising results. Random forests \citep{Breiman2001_RF} have produced competitive next-day hail predictions (Gagne et al. \citeyear{Gagne+etal2017}; \citealt{Burke+etal2019}), reliable next-day severe weather hazard guidance \citep{Loken+etal2020}, and even outperformed the Storm Prediction Center (SPC) Day 2 and 3 outlooks \citep{Hill+etal2020}. Neural networks have also shown success in predicting next-day severe weather and were more skillful than an UH baseline in \citet{Sobash+etal2020}. A key advantage of ML models is their ability to leverage multiple input predictors and learn complex relationships to produce skillful, calibrated probabilistic guidance. An additional advantage for real-time operational settings is that once an ML model has been trained, making predictions on new data is computationally quick ($\ll$ 1 s per example).

The goal of this study is to evaluate the skill and reliability of ML-based calibrations of the WoF system (WoFS) severe weather probabilistic guidance. To accomplish this goal, we trained gradient-boosted classification trees \citep{Friedman2002, Chen+Guestrin2016}, random forests, and logistic regression models on WoF System (WoFS) forecasts from the 2017-2019 Hazardous Weather Testbed Spring Forecasting Experiments (HWT-SFE; \citealt{Gallo+etal2017}) to determine which storms predicted by the WoFS will produce a tornado, severe hail, and/or severe wind report. These three ML algorithms are fairly common and have recently shown success in a variety of meteorological applications (e.g., \citealt{Mecikalski+etal2015, Erickson+etal2016}, Gagne et al.~\citeyear{Gagne+etal2017}; \citealt{Lagerquist+etal2017, Herman+Schumacher2018b, Herman+Schumacher2018a, Burke+etal2019, Loken+etal2019, McGovern+etal2019_stormlong, McGovern+etal2019_blackbox, Hill+etal2020, Jergensen+etal2020, Steinkruger+etal2020}). Recent ML studies using CAM ensemble output for severe weather prediction have only been in the next-day (24-36 hr) paradigm using grid-based frameworks (e.g., Gagne et al. \citeyear{Gagne+etal2017}, \citealt{Burke+etal2019, Loken+etal2019, Hill+etal2020, Sobash+etal2020}). Next-day forecasting methods, however, operate on a larger spatial scale because of the limited intrinsic predictability of storms at those lead times \citep{Lorenz1969} and produce overly smooth guidance compared to WoF-style forecasts, which should provide probabilistic guidance for individual thunderstorms (\citealt{Stensrud+etal2009, Stensrud+etal2013}). Therefore, we use the event-based framework based on the ensemble storm track identification method developed in \citet{Flora+etal2019}. In this framework, we can develop ML-calibrated probabilistic guidance for individual thunderstorms that produces \say{event probabilities} or the likelihood of a storm producing an event within a neighborhood determined by the ensemble forecast envelope (i.e., the forecasted uncertainty in storm location) rather than \say{spatial probabilities} or the probability of an event occurring within a prescribed radius of each model grid point (see \citealt{Flora+etal2019} for more on the distinction between event and spatial probabilities). We are also using the event-based approach since forecasters that use WoFS output focus on coherent regions of interest rather than strictly analyzing forecasts on a point-by-point basis (Wilson et al.~\citeyear{Wilson+etal2019}). 

To provide a baseline against which to test the ML models performance, the probability of 2-5 km (mid-level) UH exceeding a threshold (tuned per severe weather hazard) is extracted from each ensemble storm track similar to \citet{Flora+etal2019}.  We hypothesize that the ML-based calibrations should outperform an UH-based baseline, especially in cases of severe, non-rotating thunderstorms or in environments where supercells are less common.

% Describe the layout of the paper. 
The structure of the paper is as follows. Sections 2 and 3 describe the WoFS forecast datasets and the data processing procedures, respectively. Section 4 describes the ML models and methods used in this study. We present the results in Section 5 with conclusions and limitations of the study discussed in Section 6.
%%%%%%%%%%%%%%%%%%%%%%%%%%%%%%%
% METHODS
%%%%%%%%%%%%%%%%%%%%%%%%%%%%%%%
\section{Description of the Forecast Data}
The WoFS is an experimental multi-physics ensemble capable of producing rapidly updating severe weather guidance by frequently assimilating ongoing convection. The WoFS ensemble comprises 36 members at a 3-km horizontal grid spacing with the Advanced Research version of the Weather and Research Forecast Model (WRF-ARW; Skamarock et al. \citeyear{Skamarock+etal2008}) as the dynamic core. The physical parameterization configuration for the different ensemble members is provided in Skinner et al. (2018; their Table 1).  The initial and lateral boundary conditions for the WoFS are provided by the experimental 3-km High-Resolution Rapid Refresh Ensemble (HRRRE; Dowell et al. \citeyear{Dowel+etal2016}). The location of the WoFS domain changes daily and is centered over the region of the greatest severe weather potential. For the 2017 HWT-SFE the size of the domain was 750 x 750 km, but for subsequent HWT-SFEs is 900 x 900 km. Radial velocity, radar reflectivity, Geostationary Operational Environmental Satellite (GOES)-16 cloud water path, and Oklahoma mesonet observations (when available) are assimilated every 15 min, with conventional observations assimilated hourly. During the 2017-2018 HWT-SFEs, the ensemble adjustment Kalman filter (Anderson 2001) included in the Data Assimilation Research Testbed (DART) software was used. During the 2019 HWT-SFE, data assimilation was performed using the Community Gridpoint Statistical Interpolation based Ensemble Kalman Square Root Filter (GSI-EnKF; DTC \citeyear{DTC2017a, DTC2017b}). After five initial 15-min assimilation cycles, 18-member forecasts (a subset of the 36 analysis members) are issued every 30 min and provide forecast output every 5 min for up to 6 hours of lead time. The reader can find additional details of the WoFS in \citet{Wheatley+etal2015}, \citet{Jones+etal2016}, and \citet{Jones+etal2020}.

\begin{figure*}[t]
  \noindent\includegraphics[width=\textwidth, angle=0]{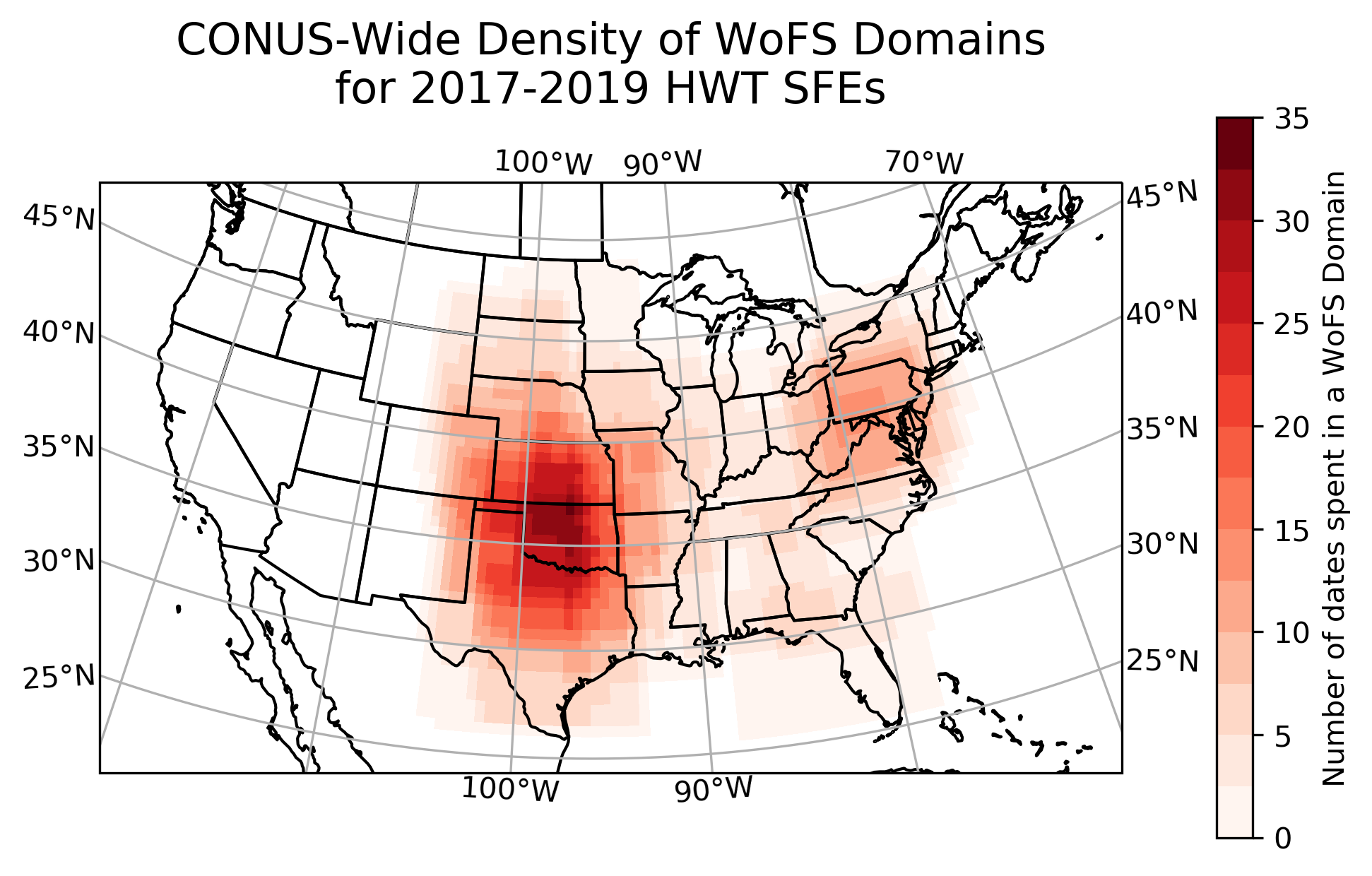}\\
  \caption{ Map of the number of times a 0.5 x 0.5 degree region was in a WoFS domain during the 2017-2019 HWT-SFEs.  }\label{fig:density_map}
\end{figure*}

This study uses 81 cases generated during the 2017-2019 HWT-SFEs. During these experiments, WoFS domains were frequently centered over the Great Plains and mid-Atlantic with less focus on the Southeast and Midwest (Fig.~\ref{fig:density_map}). This is not surprising, as severe weather is most common over the Great Plains during the spring (severe weather has a less pronounced springtime maximum over the mid-Atlantic) and becomes more common elsewhere during the summer or cool season \citep{SPC}. Overall, the dataset sufficiently samples environments relevant for springtime severe weather forecasting, but the trained ML algorithms may not be appropriate for year-round use.

To be consistent with recent WoFS verification studies (e.g., \citealt{Skinner+etal2018}) and typical National Weather Service (NWS) warning lead times \citep{Brooks+Correia2018}, the WoFS forecast data were aggregated into 30-min periods up to a lead time\footnote{It takes approximately 20–25 minutes to produce and disseminate the first two forecast hours of WoFS guidance to real-time users, so the effective lead time is shorter than the period since forecast initialization.} of 150 min (e.g., 0-30, 5-35, ..., 120-150 min).  Given the rapid model error growth on spatiotemporal scales represented in WoFS forecasts, the whole dataset was split in two based on the forecast lead time, whereby forecasts beginning in the first hour (i.e., 0-30, 5-35, ..., 60-90 min) are in one dataset (referred to as FIRST HOUR hereafter) and forecasts beginning in the second hour are in a second dataset (i.e., 65-95, 70-100, ..., 120-150 min; referred to as SECOND HOUR hereafter). The different lead times within the FIRST HOUR and SECOND HOUR are uniformly distributed (not shown). Splitting the dataset in this way allows the ML models to learn from the different forecast error characteristics in the two datasets (e.g., larger ensemble spread in SECOND HOUR than in FIRST HOUR), which should improve the models’ skill. The predictability of individual storm-scale features greatly diminishes beyond 150 min lead times \citep{Flora+etal2018}, and therefore forecasts at those lead times are not considered in this study.

\section{ Data Pre-Processing Procedures } \label{data} 
\subsection{Ensemble storm track identification and labelling}\label{sec:data} 
Object-based methods isolate important regions in a forecast space and are an effective method for reducing a large data volume into manageable components. In past ML studies using CAM ensemble output, object-based methods have been used to extract data from individual ensemble members rather than from the ensemble as a whole (e.g., Gagne et al.~\citeyear{Gagne+etal2017}, \citealt{Burke+etal2019}). However, there are limitations to extracting data from the individual ensemble members. First, applying an ML model to calibrate the individual member forecasts requires an additional procedure for combining the separate predictions into a single ensemble forecast (and potentially another round of calibration). Second, training ML models on the individual member forecasts neglects important ensemble attributes like the ensemble mean, which on average is a better prediction than any single deterministic forecast, and the ensemble spread (e.g., standard deviation), which can be a useful measure of forecast uncertainty. Past ML studies using CAM ensemble output have used ensemble statistics, but they were in a grid-based framework (e.g., \citealt{Loken+etal2020}). Therefore, we combined these past approaches by extracting ensemble information but within the event-based framework developed in \citet{Flora+etal2019}.

\begin{figure*}[t]
  \noindent\includegraphics[width=38pc,angle=0]{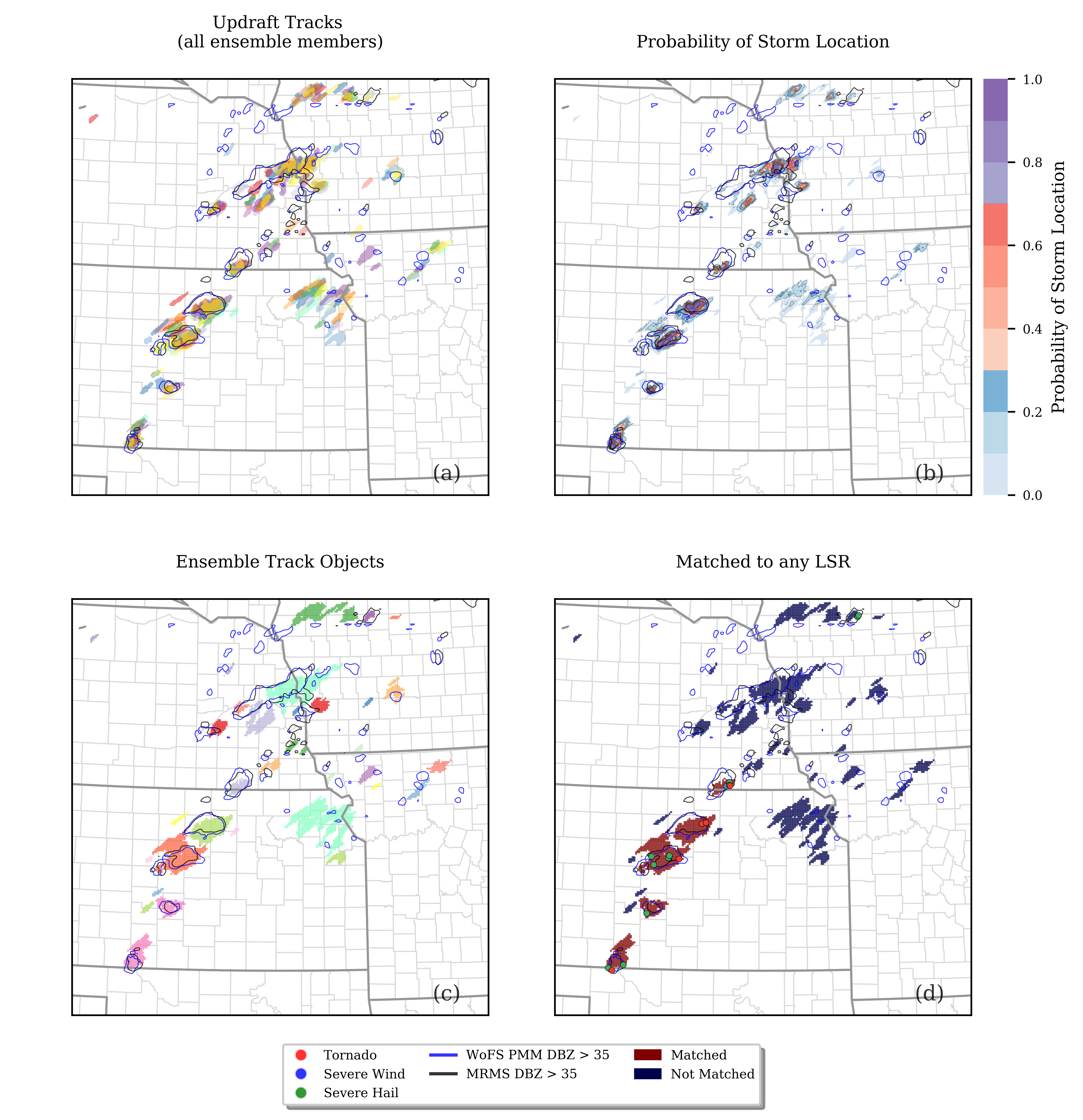}\\
  \caption{ Illustration of transforming individual ensemble member updraft tracks into ensemble storm tracks. a) Paintball plot of updraft tracks identified from 30-min-maximum column-max vertical velocity, then quality controlled as described in Section 2b.1. b) Grid-scale ensemble probability of storm location is computed from the objects in (a). c) ensemble storm track objects are identified using the algorithm outlined in Section 2b.1. d) ensemble storm track objects containing a tornado (red dot), severe hail (green dot), or severe wind (blue dot) shown in red (not matched shown in blue). The technique is demonstrated using a 0-30 min forecast initialized at 2330 UTC on 01 May 2018. For context, the 35-dBZ contour of the WoFS probability matched mean (blue) and Multi-Radar Multi-System (MRMS; black) composite reflectivity at forecast initialization time, respectively, are overlaid in each panel.  
  }\label{fig:ensemble_tracks}
\end{figure*}

An ensemble storm track, conceptually, is a region bounded by ensemble forecast uncertainty in storm location. An ensemble storm track can be composed of a single ensemble member’s storm track or some combination of up to all 18 ensemble members. Figure~\ref{fig:ensemble_tracks} shows the ensemble storm track identification procedure. First, per ensemble member, we identify storm tracks by taking peak column-maximum vertical velocity values composited over 30-min periods and thresholding them at 10 m s$^{-1}$ (Fig~\ref{fig:ensemble_tracks}a). Storm tracks not meeting a 108 km$^{2}$ (12 grid cells) minimum area threshold are removed since such storms tend to be too small and/or short-lived to be likely to produce severe weather and were found to degrade the ensemble storm track identification by producing too many objects. The ensemble probability of storm location ($EP$; Fig~\ref{fig:ensemble_tracks}b) at grid point $i$ (based on $N$ ensemble members) is calculated from the updraft tracks with the following equation:
\begin{equation}
    EP_i = \frac{1}{N}\sum_{j=1}^{N} BP_{ij}
\end{equation}
where $BP_{ij}$ (the binary probability at the $i$th grid point and $j$th ensemble member) is defined as 
\begin{equation}
 BP_{ij} = \left\{ \begin{array}{ll}
         1 & \mbox{if $i \in S_j$};\\
         0 & \mbox{if $i \notin S_j$} 
         \end{array} \right.
\end{equation}
and $S_j$ is the set of grid points within the updraft tracks for the $j$th ensemble member.  The ensemble storm track objects (Fig~\ref{fig:ensemble_tracks}c) are then identified from the $EP$ field with the following procedure:
\begin{enumerate}
    \item Apply the enhanced watershed algorithm \citep{Lakshmanan+etal2009, Gagne+etal2016} with a large area threshold (3600 km$^2$ in this study) and no minimum threshold.
    \item Apply the enhanced watershed algorithm with a smaller area threshold (2700 km$^2$ in this study) and some minimum threshold. We choose a threshold of 5.5$\%$ (one of 18 ensemble members) as setting the threshold higher than this causes excessive object break-up. 
    
    \item If an object from step 1 contains multiple objects identified in step 2, then replace the object in step 1 with those objects from step 2.
   
    \item For any remaining nonzero probabilities not assigned to an object, assign them to the closest object.
   
    \item For each grid point with a nonzero probability, assign it the object label that occurs most frequently within a 2--grid-point radius. This is necessary to quality control the previous step where points along the edge of an object can be erroneously assigned to neighboring objects.
   
    \item For objects with a solidity [ratio of object area to convex area (area of the smallest convex polygon that encloses the region)] greater than a given threshold (e.g, 1.5 in this study), revert those objects to the objects after step 2. This quality control will \say{reset} an object if the previous steps produced an object with poor solidity.
    
    \item Repeat steps 3-6 until no further changes occur. 
    
\end{enumerate}
The basis of the ensemble storm track method is the enhanced watershed algorithm which grows objects pixel-by-pixel from a set of local maxima until they reach a specified area or intensity criterion \citep{Lakshmanan+etal2009}. Objects are restricted from growing into regions where intensity falls below the prescribed minimum threshold. Once an object is identified, it restricts additional objects from growing into the region surrounding pre-existing objects to maintain object separation \citep{Lakshmanan+etal2009}. This two-pass procedure coupled with the nearest neighborhood assignment (step 5) addresses an issue raised in \citet{Flora+etal2019}: setting the enhanced watershed area threshold sufficiently low to prevent the merging of too many objects excessively reduced ensemble object size (see Fig. 3c in \citealt{Flora+etal2019}). With this improved method, the enhanced watershed may grow objects to a greater size while maintaining object separation.

After we identify the ensemble storm tracks, we classify each according to whether it contains a tornado, severe hail, and/or severe wind storm report (Fig~\ref{fig:ensemble_tracks}d). 
To account for potential reporting time errors, we considered reports within $\pm$ 15 min of either side of the 30 min forecast period (a 60 min window). Sometimes, an observed storm may produce severe weather, but there is no corresponding forecast storm in the WoFS guidance. This does not undermine the goal of the ML prediction system, which is to predict which WoFS storms will become severe. However, our inability to account for missed storm reports where the WoFS cannot predict the occurrence of a storm in a particular area highlights an important trade-off between the event-based prediction framework we use and the more traditional grid-based framework (which allows such misses to be included in the verification, but produces overly smooth forecasts).  Last, we recognize that local storm reports are error-prone (e.g., \citealt{Brooks+etal2003, Doswell+etal2005,Trapp+etal2006,Verbout+etal2006, Cintineo+etal2012, Potvin+etal2019}), but they are the best available verification database for individual severe weather hazards, have been frequently used in past ML studies (e.g., \citealt{Cintineo+etal2014, Cintineo+etal2018}, Gagne et al.~\citeyear{Gagne+etal2017}, \citealt{McGovern+etal2017,Burke+etal2019, Hill+etal2020, Lagerquist+etal2020, Sobash+etal2020, Steinkruger+etal2020}), and are used in official evaluations of NWS warnings and SPC watches and outlooks.

\subsection{Predictor Engineering}
\begin{figure*}[t]
  \noindent\includegraphics[width=38pc,angle=0]{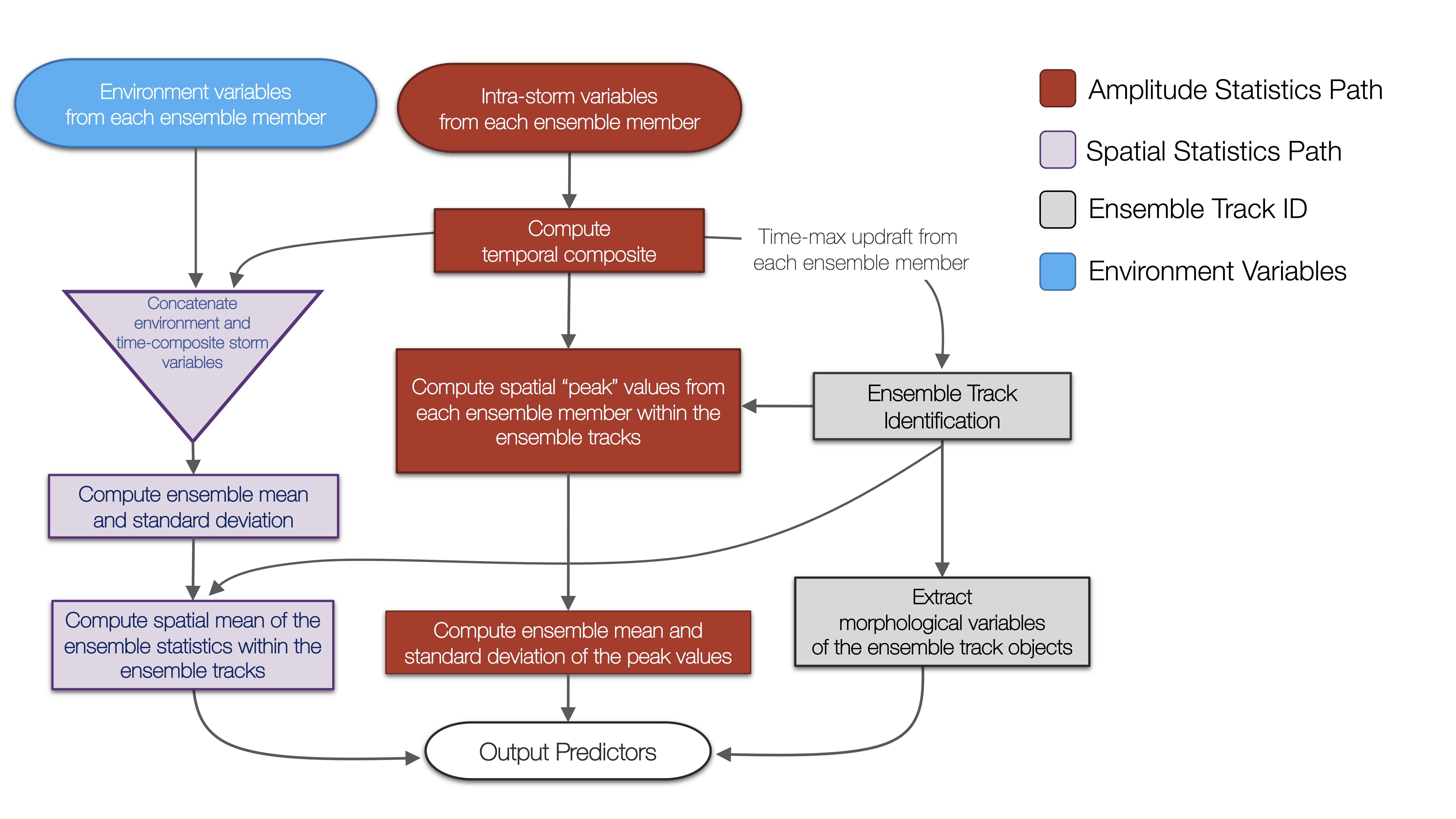}\\
  \caption{Flow chart of the data preprocessing and predictor engineering used in this study. The three components are the ensemble storm track object identification (shown in grey), the amplitude statistics (shown in red), and the spatial statistics [shown in purple (a combination of red and blue)]. Environmental variable input is shown in blue. }\label{fig:data_preprocess}
\end{figure*}

Figure \ref{fig:data_preprocess} depicts the data preprocessing and predictor engineering procedure. First, per ensemble member, the 30-min maximum (minimum) was calculated for the positively oriented (negatively oriented; denoted by $*$) intra-storm variables, and the environment variables were computed at the beginning of the valid forecast period to better sample the pre-storm environment (see Table~\ref{table:input_vars} for the input variables). 

\begin{table*}[t]
\caption{Input variables from the WoFS. The asterisk (*) refers to negatively-oriented variables. CAPE is convective available potential energy, CIN is convective inhibition, and LCL is the lifting condensation level. Mid-level lapse rate is computed over the 500-700 hPa layer and low-level lapse rate is computed over the 0-3 km layer. HAILCAST refers to maximum hail diameter from WRF-HAILCAST \citep{AdamsSelin+Ziegler2016, Adams-Selin+etal2019}. The cold pool buoyancy ($B$) is defined as $B = g \frac{\overline{\theta}_{e, z=0}}{\theta_{e,z=0}'}$ where $g$ is the acceleration due to gravity, $\overline{\theta}_{e, z=0}$ is the lowest model level average equivalent potential temperature, and $\theta_{e,z=0}'$ ($=\theta_{e,z=0} -\overline{\theta}_{e, z=0}$) is the perturbation equivalent potential temperature of the lowest model level. Values in the parentheses indicate those variables are extracted from different vertical levels or layers.} \label{table:input_vars}

    \centering
    \begin{tabular}{lll}
    \hline \hline
    Intra-storm & Environment  & Object Properties \\ 
    \hline

    Updraft Helicity (0-2 km, 2-5 km)  & Storm-Relative Helicity (0-1 km, 0-3 km) & Area\\
    Cloud Top Temperature* & 75 mb Mixed-layer CAPE  & Eccentricity\\
    0-2 km Avg. Vertical Vorticity & 75 mb Mixed-layer CIN & Orientation \\
    Composite Reflectivity & 75 mb Mixed-Layer LCL & Minor axis length \\
    1-3 km Maximum Reflectivity & 75 mb Mixed-Layer Equivalent Potential Temperature & Major axis length \\
    3-5 km Maximum Reflectivity &  U Shear (0-6 km, 0-1 km) &  Extent \\
    80-m wind speed & V Shear (0-6 km, 0-1 km) & Initialization Time \\
    10-500 m Bulk Wind Shear & 10-m U & \\
    10-m Divergence* & 10-m V & \\
    Column-maximum Updraft & Mid-Level Lapse Rate \\
    Column-minimum Downdraft* & Low-level Lapse Rate\\
    Low-level updraft (1 km AGL) & Temperature (850, 700, 500 mb) \\ 
    HAILCAST  & Dewpoint Temperature (850, 700, 500 mb) \\
    Cold Pool Buoyancy* & Geopotential Height (850, 700 500 mb) \\ 

    \hline
    \end{tabular}
\end{table*}

Predictors subsequently generated from these fields are of two modes: spatial statistics (shown as the purple path in Fig.~\ref{fig:data_preprocess}) or amplitude statistics (shown as the red path in Fig.~\ref{fig:data_preprocess}). 
For the spatial statistics, we compute the ensemble mean and standard deviation at each grid point within the ensemble storm track, then spatially average them over the storm track. We are only computing the spatial average (and not, e.g., the standard deviation within the storm track) to limit the number of predictors in favor of model interpretability over model complexity. We only compute amplitude statistics for the time-composite intra-storm variables. For the positively oriented (negatively oriented) intra-storm state variables, the spatial 90th (10th) percentile value (from grid points within an ensemble storm track) is computed from each ensemble member to produce an ensemble distribution of \say{peak} values. The 90th (10th) percentile is used as the \say{peak value} rather than maximum (minimum) since the maximum (minimum) value may be valid at only a single grid point, and therefore potentially unrepresentative. The ensemble mean and standard deviation are subsequently computed from each set of peak values to capture the expected amplitudes of storm features and the uncertainty therein. Reversing this procedure (i.e., computing the ensemble mean and standard deviation at each grid point and then finding the peak value) would have caused useful fine-scale details in the WoFS forecasts to be lost because of storm phase differences among ensemble members.

Lastly, we calculated a handful of properties describing the ensemble storm track object morphology. These include area, eccentricity, major and minor axis length, and orientation. Altogether, there are 30 amplitude statistics, 76 spatial statistics, and 7 object properties for a total of 113 predictors. 

\section{ Machine Learning Methods } \label{method} 

\subsection{Machine Learning Models} \label{sec:ml_models}
A linear regression model is a linear combination of learned weights ($\beta_i$), predictors ($x_i$) and a single bias term ($\beta_0$) :
\begin{equation}
    z = \beta_0 + \sum_{i=1}^N \beta_ix_i
\end{equation}
where $N$ is the number of predictors. For logistic regression, a logit transformation is applied to the output of the linear regression model:
\begin{equation}
    p = \frac{1}{1+\exp(-z)} 
\end{equation}
where $p$ is the model predictions [values between (0,1)]. The weights are learned by minimizing the binary cross-entropy (also known as the log-loss) between the true binary labels ($y$) and model predictions with two additional terms for regularization (known together as the elastic net penalty): 
\begin{equation}
  C \sum_{k=0}^{K}\Big[y_k\log_2(p_k) + (1-y_k)\log_2(p_k)\Big] + \frac{1-\alpha}{2}\sum_{k=0}^{K} \beta_k^2 + \alpha\sum_{k=0}^{K}|\beta_k|
\end{equation}
where $K$ is the number of training examples, C ($=\frac{1}{\lambda}$ where $\lambda \in [0, \infty)$) is the inverse of the regularization parameter (adjusts the strength of the regularization terms relative to the log-loss), and $\alpha \in [0,1]$ is a mixing parameter that adjusts the relative strength of the two regularization terms.  The second term is known as the \say{ridge} penalty or $L2$ error and it penalizes the model from heavily favoring predictors by encouraging the model to keep weights small. The last term is known as the \say{lasso} (least absolute shrinkage and selection operator) penalty or $L1$ error and it allows weights to be zeroed out thereby removing predictors from the model. Since logistic regression explicitly combines predictors (unlike the tree-based methods) and the scale of the predictors can vary considerably, we normalize each training and testing set predictor by the training dataset mean and standard deviation. We did not normalize the predictors for the tree-based methods. 

Tree-based methods are among the most common ML algorithms. A single classification tree recursively partitions a predictor space into a set of subregions using a series of decision nodes where the splitting criterion favors increasing the \say{purity} (consisting of only one class) of these regions \citep{statisticallearning}. To prevent overfitting (restricting the subregions from becoming too narrowly defined) decision trees can be \say{pruned}, for example, by requiring a maximum depth or removing final nodes (known as leaf nodes) below a minimum sample size. A classification random forest builds multiple, weakly correlated classification trees and merges their predictions to improve accuracy and stability over any individual decision tree \citep{Breiman2001_RF}. Random forests achieve the increased performance over a single decision tree by training each tree with a bootstrap resampling of the training examples and a small, random subset of predictors per split. The random forest prediction is the ensemble average of the event frequencies (from those examples in the leaf node) predicted by each individual classification tree (all trees are weighed equally). In contrast, an ensemble of decision trees can be combined using the statistical method known as gradient boosting where predictions are not made independently, but sequentially \citep{Friedman2002}. The first tree is trained on the true targets, and then each additional tree is trained on the error residual of the previous tree. Conceptually, trees are added one at a time with each successive tree structure adjusted based on the results of the previous iteration. The final prediction of a gradient-boosted forest is the weighted sum of the predictions from the separate classification trees.

ML models may correctly rank predictions (predict the most probable class), yet produce highly uncalibrated probabilistic output, especially when trained on data in which the ratio of events to non-events departs substantially from the climatology event frequency. Isotonic regression is a non-parametric method for finding a non-decreasing (monotonic) approximation of a function and is commonly used for calibrating ML predictions \citep{Niculescu-Mizil2005}. Past studies in weather-based studies have found success using isotonic regression-based calibrations \citep{Lagerquist+etal2017,McGovern+etal2019_stormlong, Burke+etal2019}. To compute calibrated probability estimates, isotonic regression seeks the best fit of the data that are consistent with the classifier’s ranking. First, pairs of $(p_i,y_i)$ are sorted based on $p_i$ where $p$ is the base classifier’s uncalibrated predictions and $y$ is the true binary labels. Starting with $y_1$, the algorithm moves to the right until it encounters a ranking violation $(y_i > y_{i+1}; 0 > 1)$. Pairs $(y_i, y_{i+1})$ with ranking violations are replaced by their average and potentially averaged with previous points to maintain the monotonicity constraint. This process is repeated until all pairs are evaluated. The outcome is a model that relates a base classifier’s prediction to a calibrated conditional event frequency (through the averaging of the rank violations). To prevent introducing bias, the isotonic regression is typically trained on the predictions and labels of the base model on a validation dataset. Rather than training on an independent validation dataset, we use the cross-validation approach from \citet{Platt99} where the base model is fit on each training fold and used to make predictions on the corresponding validation fold. The calibration model (e.g., isotonic regression) is then trained on the concatenation of the predictions from the different cross-validation folds. The base model can then be refit to the whole training dataset, while the calibration model is effectively fit on the whole dataset without biasing the predictions.

In this study, we are using the random forest and logistic regression models available in the sci-kit learn package \citep{Pedregosa+etal2011}. The gradient-boosted classification trees (XGBoost hereafter) model comes from the open-source eXtreme Gradient Boosted (XGBoost) package \citep{Chen+Guestrin2016}. The calibration model used is the isotonic regression model available in the sci-kit learn package \citep{Pedregosa+etal2011}.

%%%%%%%%%%%%%
% BASELINE 
%%%%%%%%%%%%%

\subsection{Developing a Baseline Prediction from the WoFS}
The baseline prediction is the ensemble probability of mid-level UH exceeding a threshold, given the prior success of this diagnostic in predicting severe weather and its frequent use as a baseline in other severe-weather-based ML studies (e.g., \citealt{Gagne+etal2017, Loken+etal2020, Sobash+etal2020}). The ensemble probabilities are computed using equation (1), but the binary probability for the $j$th ensemble member at the $i$th grid point is defined as
\begin{equation}
 BP_{ij} = \left\{ \begin{array}{ll}
         1 & \mbox{if UH$_{ij}$ $\geq$ t};\\
         0 & \mbox{if UH$_{ij}$ $<$ t} 
         \end{array} \right.
\end{equation}
where $t$ is the UH threshold. We then set the event probability for a storm to the maximum ensemble probability within the ensemble storm track, similar to the method used in \citet{Flora+etal2019}. 
\begin{figure*}[t]
  \noindent\includegraphics[width=38pc,angle=0]{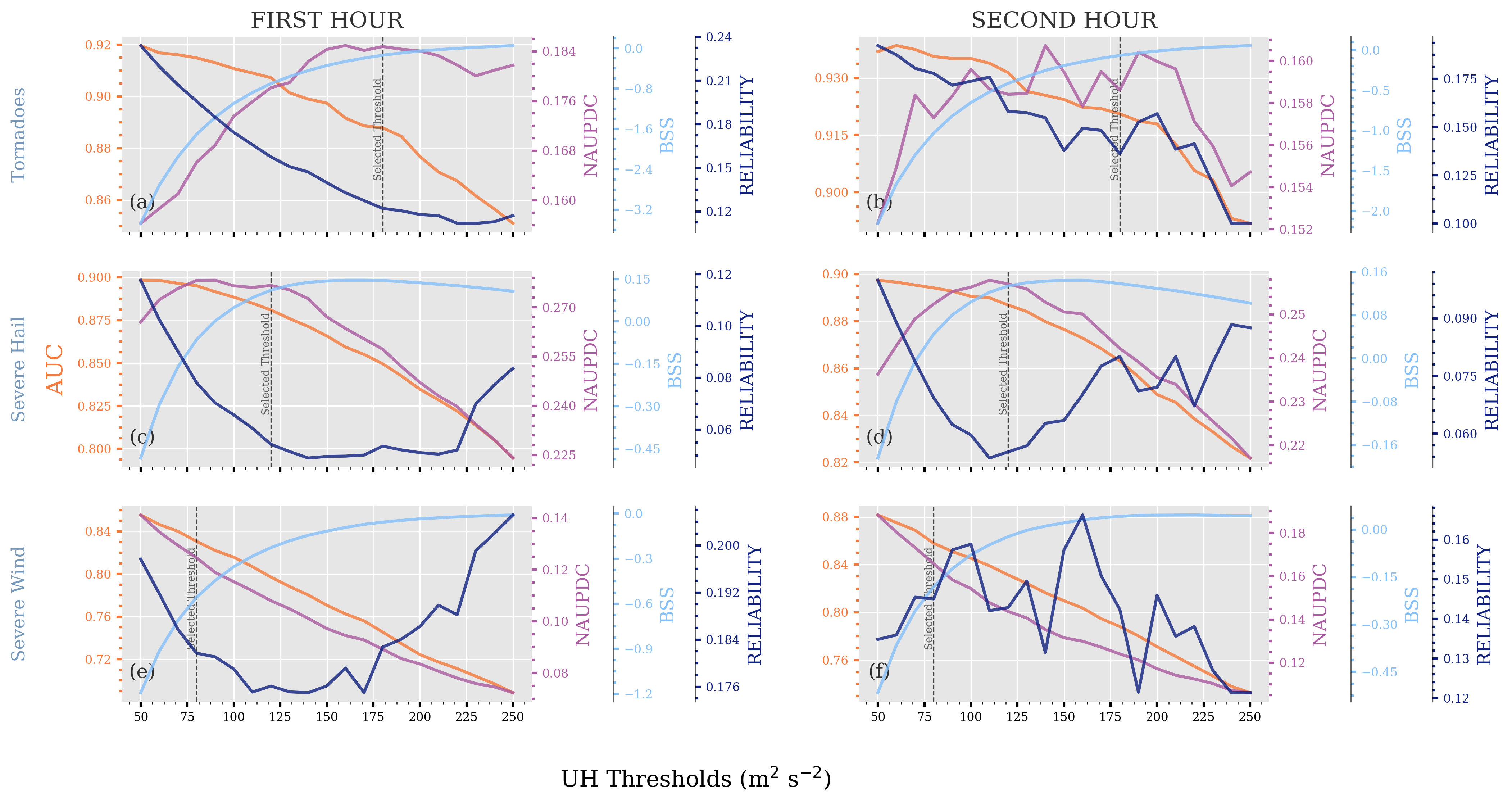}\\
  \caption{Cross-validation average (within the training dataset) performance of the baseline updraft helicity probabilities as a function of a varying threshold for predicting tornadoes (top row), severe hail (middle row), and severe wind (bottom row). Panels on the left (right) are valid for FIRST HOUR (SECOND HOUR). Metrics include AUC (orange), Normalized AUPDC (NAUPDC; purple), Brier skill score (BSS; light blue), and the reliability component of the BSS (RELIABILITY; dark blue). The vertical dashed line labelled 'selected threshold' indicates the updraft helicity threshold which optimizes certain metrics or limits tradeoffs between the various metrics (see text for details).}\label{fig:uh_threshs}
\end{figure*}
To tune the threshold for each severe weather hazard, we tested the mid-level UH probabilities on the 5 validation folds (described above) and computed the cross-validation average performance for multiple metrics (Fig.~\ref{fig:uh_threshs}). Changing the UH threshold reveals there is a tradeoff between the ranking-based and calibration-based metrics (defined in section 5).  Increasing the threshold improves reliability, but decreases the ability of the probabilities to discriminate between events and non-events.  For FIRST HOUR tornado prediction, we selected a threshold of UH $>$ 180 m$^2$ s$^{-2}$  since a higher threshold degrades the ranking-based metrics although reliability continues to improve (Fig.~\ref{fig:uh_threshs}a). A similar argument can be made for the 120 m$^2$ s$^{-2}$ threshold selected for severe hail (Fig.~\ref{fig:uh_threshs}b). The higher threshold for tornadoes than severe hail suggests that the ensemble is discriminating between strong and weak rotation, contrary to results of \citet{Sobash+etal2016}, which found that higher mid-level UH thresholds had poor discrimination. For severe wind (Fig. \ref{fig:uh_threshs}e), there is no apparent optimal threshold, suggesting the UH is not the most appropriate predictor of severe wind likelihood. As a compromise, we choose a threshold of UH $>$ 80 m$^2$ s$^{-2}$ to be consistent with \citet{Flora+etal2019}. The results are similar in the SECOND HOUR dataset and therefore we kept the optimal threshold the same for simplicity (Fig. \ref{fig:uh_threshs}b, d, f).

\subsection{Model Tuning and Evaluation}
\begin{figure*}[t]
  \noindent\includegraphics[width=38pc,angle=0]{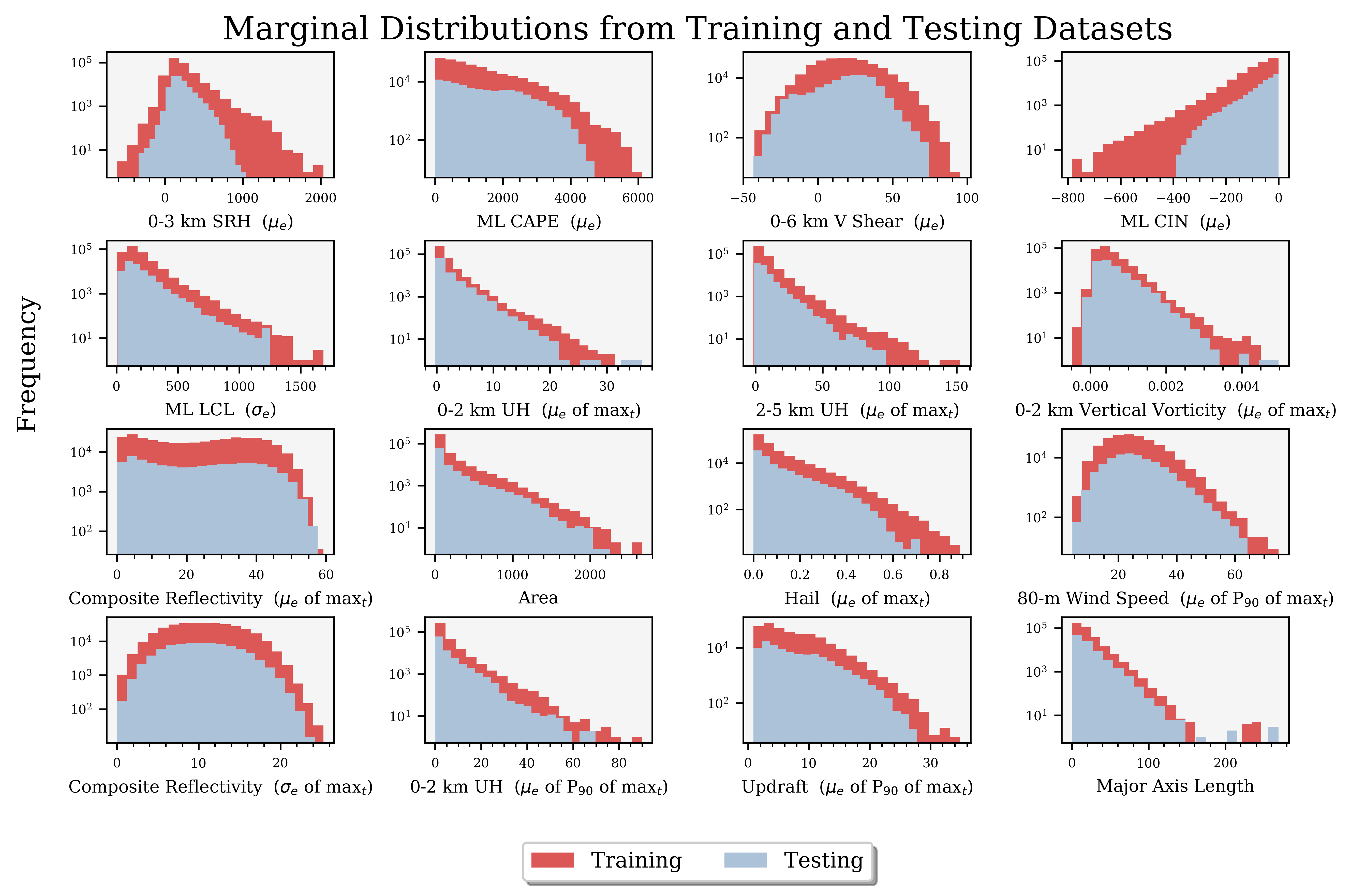}\\
  \caption{ Marginal distributions of the training and testing dataset for a subset of predictors setup for predicting tornadoes in the FIRST HOUR dataset. ($\mu_e$) refers to spatial-average ensemble mean of the environmental variables, ($\mu_e$ of max$_t$) is spatial-average ensemble mean of the time-composite intra-storm variables, ($\mu_e$ of $P_{90}$ of max$_t$) is the ensemble-average of the spatial 90th percentile values extracted from ensemble members within the ensemble storm tracks, and ($\sigma_e$ of max$_t$) is spatial-average ensemble standard deviation of the time-composite intra-storm variables. SRH is the storm-relative helicity, and Hail refers to maximum hail diameter from WRF-HAILCAST.}\label{fig:distributions}
\end{figure*}

To assess expected model performance, both the FIRST HOUR and SECOND HOUR datasets were split into 65 dates for training and 16 dates for testing, respectively. Rather than randomly separating the dates, we ensured that the ratio of dates with at least one event to the total number of dates was maintained for both the training and testing partitions.  For example, if 40 of the 81 dates had a tornado (50$\%$), then this ratio was approximately maintained in both the training and testing dataset.  This simple approach helps ensure that the testing dataset is more representative of the training dataset, which limits bias in the assessment of model performance. We provide the number of examples in each training and testing dataset per hazard in Table \ref{table:num_of_examples}. 
\begin{table}[t]
\caption{Numbers of examples in the training and testing datasets for the different severe weather hazards and lead time intervals.} \label{table:num_of_examples}
\begin{center}
\begin{tabular}{llcc}
\hline \hline
                   &    & Training & Testing  \\
\hline                 
FIRST HOUR & & &  \\
              & Tornado     & 346 341 & 82 750 \\
              & Severe Hail & 349 508 & 79 583  \\ 
              & Severe Wind & 330 840 & 98 251 \\
SECOND HOUR & & &  \\
              & Tornado     & 262 878 & 82 483 \\
              & Severe Hail & 258 270 & 87 091  \\ 
              & Severe Wind & 258 991 & 86 370 \\
\hline
\end{tabular}
\end{center}
\end{table}
Figure \ref{fig:distributions} shows the marginal distribution for select predictors from the training and testing dataset for FIRST HOUR tornado predictions. The overall distribution of training and testing sets are similar for a majority of the predictors (results were similar for the other hazards and SECOND HOUR dataset; not shown). The distribution of these predictors in the training and testing datasets also had considerable overlap for examples only from the positive class (matched to an LSR).

Bayesian hyperparameter optimization (hyperopt; \citealt{Bergstra2013}) was used to identify the optimal hyperparameters for each model using 5-fold cross validation over the training dataset. The hyperopt python package is based on a random search method but implements a Bayesian approach where performance on previous iterations helps determine the optimal parameters. For this study, we are using the area under the performance diagram curve (defined in section 5\ref{ssec:perform_diagrams}) as our optimization metric. The default stopping criterion in hyperopt is a user-set maximum number of evaluation rounds, so we implemented an early stopping criterion where a 1$\%$ improvement in performance must occur within a set amount of rounds or else optimizing stops, which improves computational efficiency (we found that requiring said improvement at least every 10 rounds was sufficient). The hyperparameters and values used for each model are presented in Table \ref{table:hyperparameters_values}. 
\begin{table*}[t]
\caption{Hyperparameter values attempted for each model in the hyperparameter optimization.}\label{table:hyperparameters_values}
\begin{center}
\begin{tabular}{llcc}
\hline \hline
                    & Hyperparameter & Values  \\
\hline                 
Random Forest & & &  \\
              & Num. of Trees  & 100,250,300,500, 750, 1000, 1250, 1500 \\
              & Maximum Depth & 5, 10, 15, 20, 30, 40, None  \\ 
              & Minimum Leaf Node Sample Size & 1, 5, 10, 15, 25, 50 \\
XGBoost & & &  \\
             & Num. of Trees & 100,250,300,500, 750, 1000, 1250, 1500 \\
             & Minimum loss reduction ($\gamma$)    & 0, 0.001, 0.01, 0.3, 0.5, 1  \\ 
             & Maximum Depth & 2,4,7,10 \\
             & Learning Rate ($\eta$)  & 10$^{-1}$, 10$^{-2}$, 10$^{-3}$, 10$^{-4}$ \\
             & Minimum Child Weight & 1, 5, 10, 15, 25 \\ 
             & Ratio of predictors randomly selected per tree & 0.7, 0.8, 1.0 \\
             & Subsample ratio of the examples  & 0.5, 0.6, 0.7, 1.0 \\
             & $L_1$ weight & 0, 0.5, 1, 10, 15  \\ 
             & $L_2$ weight & 0.0001, 0.0005, 0.001, 0.005, 0.01, 0.1, 1.0 \\
Logistic Regression & & &  \\
              & C & 0.0001, 0.001, 0.01, 0.1, 1.0 \\
              & $\rho$ (l1$\_$ratio) & 0.0001, 0.001, 0.01, 0.5, 1.0  \\ 

\hline
\end{tabular}
\end{center}
\end{table*}
For those hyperparameters not listed, we used the default values in version 0.22 of the scikit-learn software \citep{Pedregosa+etal2011} and version 0.82 of the XGBoost software \citep{Chen+Guestrin2016}. The optimal hyperparameter values for each model and severe weather hazard for the FIRST HOUR and SECOND HOUR dataset are provided in Table~\ref{table:parameters_first_hour} and Table~\ref{table:parameters_second_hours}, respectively.
\begin{table*}[t]
\caption{Optimal hyperparameter values for each model and severe weather hazard for the FIRST HOUR dataset.} \label{table:parameters_first_hour}
\begin{center}
\begin{tabular}{llccc}
\hline \hline
                    & Hypermeter & Tornadoes & Severe hail & Severe Wind  \\
\hline                 
Random Forest & & & & \\
              & Num. of Trees                 & 100 & 1500 & 250  \\
              & Maximum Depth                 & 40  & 40   & 20   \\ 
              & Minimum Leaf Node Sample Size & 10  & 1    & 1  \\
XGBoost & & & & \\
             & Num. of Trees                                  & 300 & 250 & 300 \\
             & Minimum loss reduction ($\gamma$)              & 0.5 & 0   & 0   \\ 
             & Maximum Depth                                  & 10 & 10  & 7  \\
             & Learning Rate ($\eta$)                         & 0.1 & 0.1 & 0.1  \\
             & Minimum Child Weight                           & 1 & 1 & 15  \\ 
             & Ratio of predictors randomly selected per tree & 0.7 & 0.8 & 0.8  \\
             & Subsample ratio of the examples                & 1.0 & 0.6 & 1.0  \\
             & $L_1$ weight ($\alpha$)                        & 0.5 & 1 & 1   \\ 
             & $L_2$ weight ($\lambda$)                       & 0.001 & 0.0005 & 0.1  \\
Logistic Regression & & & &  \\
              & C & 0.1 & 0.01 & 0.01  \\
              & $\rho$ (l1$\_$ratio) & 0.0001 & 0.01 & 0.001   \\

\hline
\end{tabular}
\end{center}
\end{table*}

\begin{table*}[t]
\caption{Same as in Table \ref{table:parameters_first_hour}, but the SECOND HOUR dataset.} \label{table:parameters_second_hours}
\begin{center}
\begin{tabular}{llccc}
\hline \hline
                    & Hypermeter & Tornadoes & Severe hail & Severe Wind  \\
\hline                 
Random Forest & & & & \\
              & Num. of Trees                 & 1250 & 1250 & 250  \\
              & Maximum Depth                 & 20  & 20   & 40   \\ 
              & Minimum Leaf Node Sample Size & 50  & 5    & 5  \\
XGBoost & & & & \\
             & Num. of Trees                                  & 250 & 500 & 300 \\
             & Minimum loss reduction ($\gamma$)              & 0 & 0   & 1.0   \\ 
             & Maximum Depth                                  & 10 & 10  & 10  \\
             & Learning Rate ($\eta$)                         & 0.1 & 0.1 & 0.1  \\
             & Minimum Child Weight                           & 10 & 5 & 25  \\ 
             & Ratio of predictors randomly selected per tree & 0.7 & 1.0 & 0.8  \\
             & Subsample ratio of the examples                & 0.7 & 1.0 & 0.7 \\
             & $L_1$ weight ($\alpha$)                        & 1 & 0.5 & 10   \\ 
             & $L_2$ weight ($\lambda$)                       & 0.01 & 0.1 & 1.0  \\
Logistic Regression & & & &  \\
              & C & 0.01 & 0.01 & 0.01  \\
              & $\rho$ (l1$\_$ratio) & 0.001 & 1.0 & 1.0   \\ 

\hline
\end{tabular}
\end{center}
\end{table*}

For the final assessment, we evaluated the ML models and UH-based baselines on the independent testing datasets (severe weather hazard dependent). All metrics are bootstrap resampled (N=1000) to produce confidence intervals for significance testing. For an unbiased measure of variance, the bootstrapping method requires independent samples, but our testing samples come from overlapping forecast ranges (e.g., 0-30, 5-35, 10-40, etc.) and therefore are not independent. We do not track the ensemble object in time  (and therefore we cannot compute serial correlations on the full dataset), but based on a manual analysis of a small subset, we found that serial correlations for some predictors were not negligible (e.g., r=0.2), but small enough that the confidence intervals should not markedly underestimate the true uncertainty of the various verification scores. Lastly, the following verification results are aggregated over each dataset, FIRST HOUR and SECOND HOUR, respectively, but we found that performance is fairly consistent (with some variance) at each lead time (not shown).

%%%%%%%%%%%%%%%%%%%%%%%%%%%%%%%%%%%%%%%%%%%%%%%%%%%%
% RESULTS SECTION 
%%%%%%%%%%%%%%%%%%%%%%%%%%%%%%%%%%%%%%%%%%%%%%%%%%%%
\section{Results }\label{sec:results}
The verification methods for this study include the receiver operating characteristic (ROC) curve \citep{Metz1978}, performance diagram \citep{Roebber2009}, and the attribute diagram \citep{Hsu+Murphy1986}. The ROC curve and performance diagram are derived from converting forecast probabilities to a set of yes/no forecasts based on different probability thresholds and computing contingency table metrics. The four components of the contingency table are
\begin{enumerate}
    \item \say{hits}: forecast \say{yes} for a given hazard and the ensemble storm track is matched to a corresponding LSR
    \item \say{misses}: forecast \say{no} for a given hazard, but the ensemble storm track is matched to a corresponding LSR
    \item \say{false alarms}: forecast \say{yes} for a given hazard, but the ensemble storm track is not matched to a corresponding LSR
    \item \say{correct negatives}: forecast \say{no} for a given hazard and the ensemble storm track is not matched to a corresponding LSR
\end{enumerate}
The most common contingency metrics include probability of detection (POD; $\frac{a}{a+c}$), probability of false detection (POFD; $\frac{b}{b+d}$), success ratio (SR; $\frac{a}{a+b}$), false alarm ratio (FAR; $\frac{b}{a+b}$), critical success index (CSI; $\frac{a}{a+b+c}$), and frequency bias ($\frac{a+b}{a+c}$) where $a,b,c,d$ are the number of hits, false alarms, misses, and correct negatives, respectively.

\subsection{Sensitivity to Class Imbalance}\label{sec:class_imb}
The full dataset (combined FIRST HOUR and SECOND HOUR) used in this study is heavily imbalanced towards non-events; 1.2$\%$, 2.5$\%$, and 4$\%$ of ensemble storm track objects are matched to a tornado, severe hail, or severe wind report, respectively. ML algorithms often struggle to learn patterns and relationships from imbalanced datasets \citep{Batista+etal2004, Sun+etal2009}. One method to counteract the class imbalance is to randomly undersample the majority class (i.e., non-events) to produce a balance of events and non-events.  For all three ML algorithms, randomly undersampling the majority class significantly improved tornado prediction as compared to training on the original dataset. However, for severe wind and hail, the difference in performance for all three ML algorithms training on resample data versus the original training dataset was negligible. We propose two reasons for this result. First, a significant number of ensemble storm tracks are small (e.g., only composed of a single ensemble member’s updraft track) and these are rarely matched to storm reports, making them easily distinguishable as non-events. Thus, given that the effective ratio of events to non-events is likely much higher for severe wind and hail, the class separation may be large enough to compensate for the class imbalance. Second, tornadoes are much rarer than the other two hazards and our understanding of the processes and environmental characteristic separating tornadic and non-tornadic environments remains an active area of research (e.g., \citealt{Anderson-Frey+etal2017,Coffer+etal2017,Coffer+etal2019,Coniglio+Parker2020, Flournoy+etal2020}). For example, \citet{Coffer+etal2017} and \citet{Flournoy+etal2020} found that chaotic intra-storm processes (which may not be resolved on a 3-km grid) can lead to weak tornadoes in environments that are otherwise characterized as non-tornadic. Therefore, eliminating a large portion of non-events (which can be associated with missing reports) from the training dataset improves the signal-to-noise ratio more for tornadoes than for severe wind and hail. 

\subsection{Example Forecasts}\label{ssec:example_forecasts}
\begin{figure*}[t]
  \noindent\includegraphics[width=38pc,angle=0]{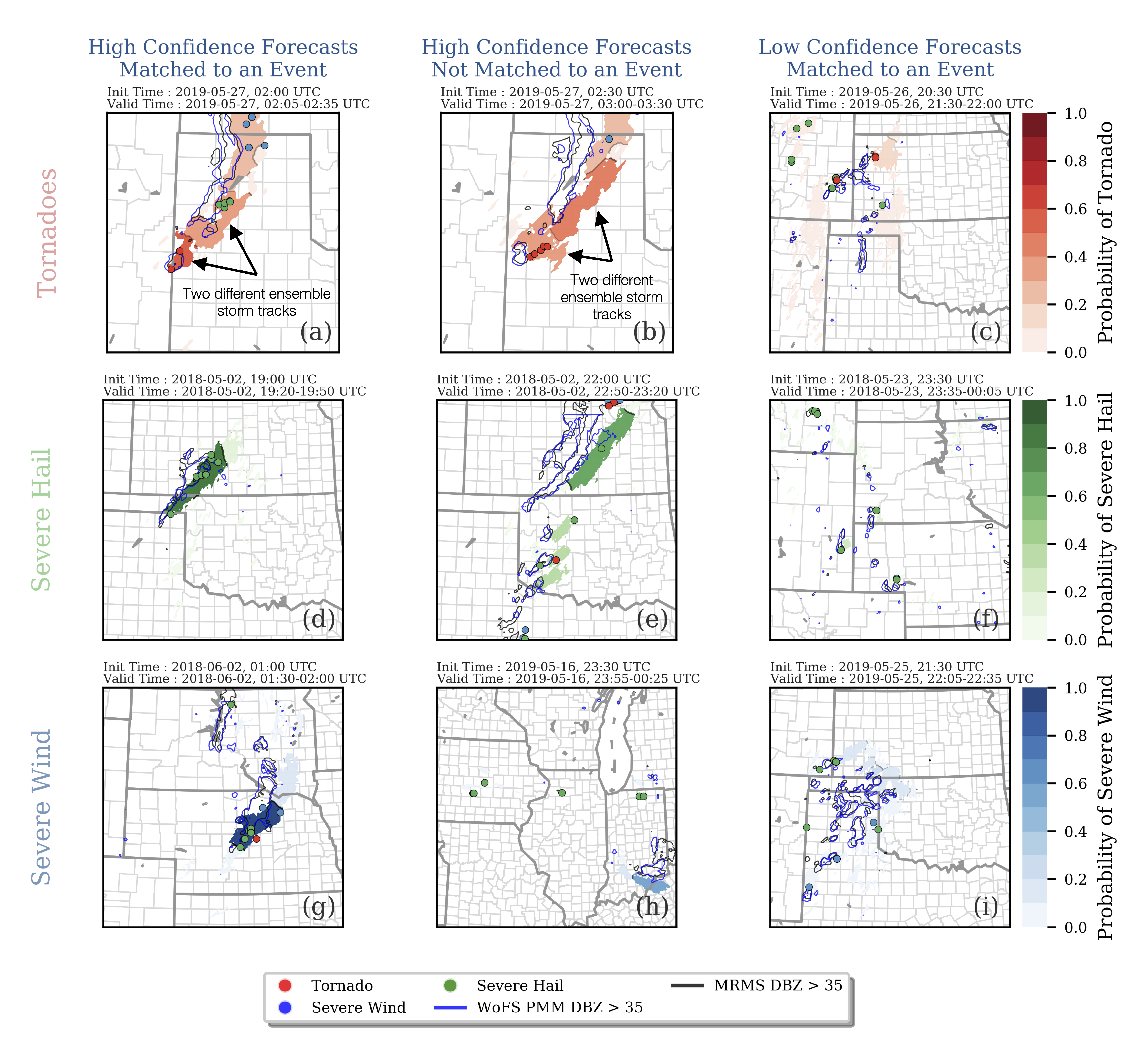}\\
  \caption{ Examples forecast from the random forest model predicting tornadoes (first row), severe hail (middle row), and severe wind (bottom row). These forecasts are representative instances of (first column) a high confidence forecast matched to an event (middle column) a high confidence forecast not matched to an event and (last column) a low confidence forecast matched to an event. For context, the 35-dBZ contour of the WoFS probability matched mean (blue) and Multi-Radar Multi-System (MRMS; black) composite reflectivity at forecast initialization time, respectively, are overlaid in each panel. The forecast initialization and valid forecast period are provided in the upper left hand corner of each panel. Tornado, severe hail, and severe wind reports are shown as red, green, and blues circles, respectively. The tornado forecasts in panel (a) and (b) have been zoomed in to focus on the isolated supercell and the southern end of the MCS over the TX Panhandle.  The annotation highlights the two different ensemble storm tracks associated with the MRMS convection.}\label{fig:example_forecast}
\end{figure*}

Figure~\ref{fig:example_forecast} shows characteristic examples of good and poor forecasts from the random forest model; these represent the other models as well (not shown). 
These examples include high confidence (probabilities closest to 1) forecasts matched and not matched to an event and low confidence (probabilities closest to 0) forecasts matched to an event. The skill of the ML forecasts is largely driven by the ability of the WoFS to accurately analyze ongoing convection through data assimilation. The classification, however, as we will see, is sensitive to slight changes in object location/separation. There may be minimal subjective differences between a confident match and confident false alarm (high confidence forecast not matched to the event), which is a limitation of the current method. For example, for high confidence (higher probabilities) forecasts matched to an event, the convection is fairly organized, and the WoFS matches well with the observed reflectivity (Fig ~\ref{fig:example_forecast}a,d,g). Unfortunately, high confidence forecasts not matched to an event can exhibit similar behavior (Fig ~\ref{fig:example_forecast}b,e,h). In Fig.~\ref{fig:example_forecast}a and Fig.~\ref{fig:example_forecast}b, storms in the Texas Panhandle have similar tornado probabilities despite only one of them producing tornado LSRs. It is possible that in this case the useful information for tornado forecasting in the WoFS was confined to larger spatial scales preventing discrimination of tornadic and non-tornadic storms occurring in proximity to one another. Complicating the interpretation, some of these apparent forecast busts may in fact be associated with an unreported event.For example, \citet{Potvin+etal2019} found that over 50$\%$ of tornadoes went unreported from 1975 to 2016. For severe wind (Fig.~\ref{fig:example_forecast}h), the timing of the higher confidence forecast was early as severe wind reports were eventually observed on the border of southern Ohio and northwest Kentucky (though the observed storms were outside the WoFS domain).

For low confidence forecasts of severe hail and severe wind matched to an event, the convection is discrete and poorly organized (Fig ~\ref{fig:example_forecast}f ) or disorganized and complex (Fig ~\ref{fig:example_forecast}i). For the first case, discrete, poorly organized convection suggests a weakly forced environment that has lower predictability and in which it is more difficult to produce an accurate ensemble analysis.  For the second case the WoFS reflectivity generally agrees with the observed reflectivity well, but the severe wind reports are associated with the weaker, isolated convection, which can have limited predictability as well (similar for tornadoes; Fig.~\ref{fig:example_forecast}c).

LSRs sometimes occur just outside of the boundaries of the ensemble storm tracks; see, for example, the severe hail report associated with the northernmost storm in Oklahoma in Fig ~\ref{fig:example_forecast}e. On the other hand, the ensemble storm track areas are larger than a typical warning polygon and represent the WoFS’s full range of storm location, and so our matching criterion is already relatively lenient. Given the impact of misses arising from small spatial errors in forecast storm tracks and spurious false alarms arising from missing reports, however, we argue that the following verification results likely underestimate the true skill of the ML models.

\subsection{ROC Diagrams}\label{ssec:roc_diagrams}
\begin{figure*}[t]
  \noindent\includegraphics[width=38pc,angle=0]{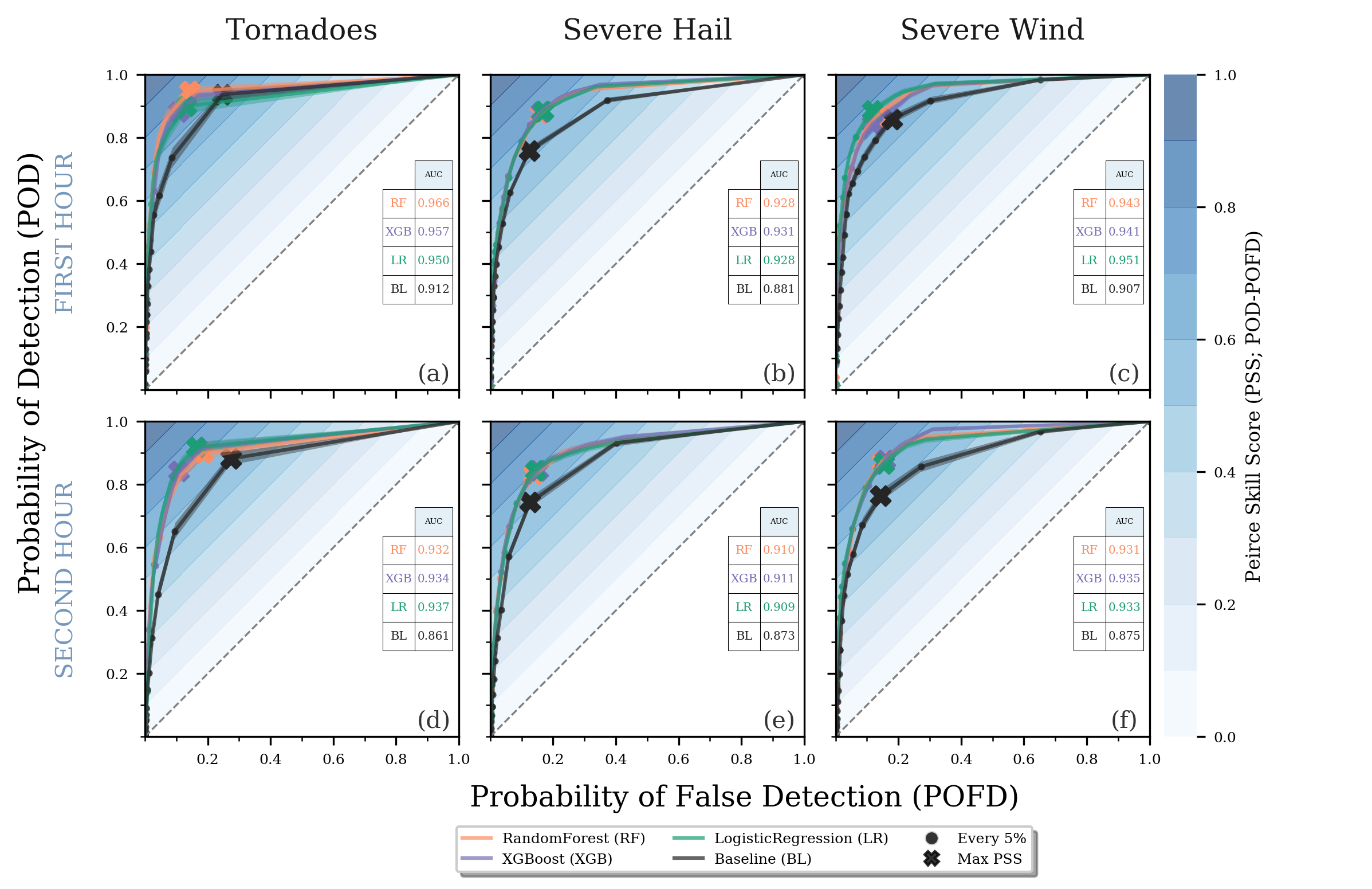}\\
  \caption{ROC curves for the random forests (RF;light orange), gradient-boosted classifier trees [XGBoost(XGB); light purple], logistic regression (LR;green), and UH baseline (BL; black) predicting whether an ensemble storm track will contain a tornado (first column), severe hail (second column), or severe wind (third column) report. Results are combined over 30-min predictions starting within the lead times in the first hour (i.e., 0-30, 5-35, ..., 60-90 min; shown in panels a, b, c) and in the second hour (i.e., 65-95, 70-100, ..., 120-150 min; shown in panels d,e,f), respectively. Each line (shaded area) is the mean (95$\%$ confidence interval), determined by bootstrapping the testing examples (N=1000). Curves were calculated every 0.5$\%$ with dots plotted every 5$\%$. The diagonal dashed line indicates a random classifier (no-skill). The mean AUC for each model is provided in the table in the upper right hand side of each panel. 
  The filled contours are the Pierce skill score (PSS; also known as the true skill score) which is defined as POD-POFD. The maximum PSS is denoted on each curve with an X.  }\label{fig:roc_curves}
\end{figure*}

% ROC DIAGRAMS
The ROC curve plots POD against POFD for a series of probability thresholds and, coupled with the area under the ROC curve (AUC), assesses the ability of the forecast system to discriminate between events and non-events. An AUC = 0.5 indicates a no-skill prediction while a perfect discriminator will score an AUC=1.  The ROC curve results are shown in Figure \ref{fig:roc_curves}. 
All three ML models produced, on average, an AUC greater than 0.9 for all three severe weather hazards for both lead time sets. While the ML model AUC scores were significantly better than those for the UH baseline, the latter were near or above 0.9, suggesting that the WoFS UH guidance is already a fairly good discriminator for the three severe weather hazards. While the AUC is high, it’s important to consider that this score is invariant to class imbalance and weighs event and non-event examples equally. Thus, the AUC provides an overly optimistic assessment of discrimination in applications where less importance is placed on correctly predicting non-events. For severe weather prediction, correct negatives are conditionally important because it is only desirable to accurately predict non-events in environments that favor severe weather (to reduce false alarms).  However, a large number of ensemble storm tracks are easily distinguishable as non-events (as mentioned in section 4\ref{sec:class_imb}), which suggests that caution be exercised when interpreting the high AUC values in this study. This effect also explains why AUC increases for severe weather hazards with lower climatological event frequencies; for rarer events, the aforementioned ensemble storm tracks become even easier to identify as non-events.

\subsection{Performance Diagrams}\label{ssec:perform_diagrams}
The performance diagram\footnote{Commonly known as the precision-recall diagram \citep{Manning+Schutze1999} in the ML community where recall is POD and precision is SR} plots the SR against the POD for a series of probability thresholds and assesses the ability of the model to correctly predict an event while ignoring correct negatives \citep{Roebber2009}. The performance diagram is complementary to the ROC curve, especially for imbalanced prediction problems (like severe weather forecasting) where it is more important to correctly predict events than non-events \citep{Davis+Goadrich2013}. CSI and frequency bias are functionally related to POD and SR and are also displayed on the performance diagram. A probabilistic forecast is considered to have perfect performance when the CSI and frequency bias are equal to 1 (corresponding to the upper right corner) for some probability threshold. However, for probabilistic forecasts of rare events, a maximum CSI of 1 is practically unachievable \citep{Hitchens+etal2013} and the maximum CSI tends to be associated with a frequency bias $>$ 1 \citep{Baldwin+Kain2006}. 

Similar to the ROC Diagram, one can compute the area under the performance diagram curve (AUPDC\footnote{Also known as the area under the precision-recall curve, which is often acronymized as AUPRC or AUCPR}). Rather than computing the area through integration, which can be too optimistic, it is more robust to compute AUPDC from the weighted average of SR\footnote{Known better by the term \say{average precision} where precision is synonymous with success ratio} \citep{Boyd+etal2013}: 
\begin{equation}
    AUPDC = \sum_{k=1}^K (POD_k - POD_{k-1})SR_k
\end{equation}
where $K$ is the number of probability thresholds used to calculate POD and SR. Unlike AUC, AUPDC is not invariant to class imbalance because the number of possible false alarms is dependent on the class balance and changing the ratio of events to non-events will affect the minimum possible SR. The minimum SR was defined in \citet{Boyd+etal2012} as
\begin{equation}\label{eqn:sr_min}
    SR_{min} = \frac{\pi POD}{1 - \pi + \pi POD}
\end{equation}
where $\pi$ is the climatological event frequency of the dataset (number of events divided by the total number of examples). If a curve lies along $SR_{min}$, the prediction system is considered to have no skill. Therefore, one can normalize AUDPC by the minimum possible AUPDC \citep{Boyd+etal2012}, which facilitates comparing the model skill on datasets with different climatological event frequencies for a given hazard or comparing model performance for different hazards with different climatological event frequencies. The minimum AUPDC is:
\begin{equation}
    AUPDC_{min} = \frac{1}{pos} \sum_{i=1}^{pos}\frac{i}{i+neg}
\end{equation}
where $pos$ and $neg$ are the number of event and non-event examples in the testing dataset, respectively \citep{Boyd+etal2012}.
The normalized AUPDC (NAUPDC) is defined as:
\begin{equation}
    NAUPDC = \frac{AUPDC - AUPDC_{min}}{1 - AUPDC_{min}}
\end{equation}
Regardless of climatological event frequency, the best possible classifier will have an NAUPDC of 1 and the worst possible classifier will have an NAUPDC of 0. We can also normalize the maximum CSI by the maximum CSI of a no-skill system [equal to the climatological event frequency ($\pi$); derivation provided in the appendix] using a computation similar to equation 10 (hereafter referred to as NCSI). 

\begin{figure*}[t]
  \noindent\includegraphics[width=38pc,angle=0]{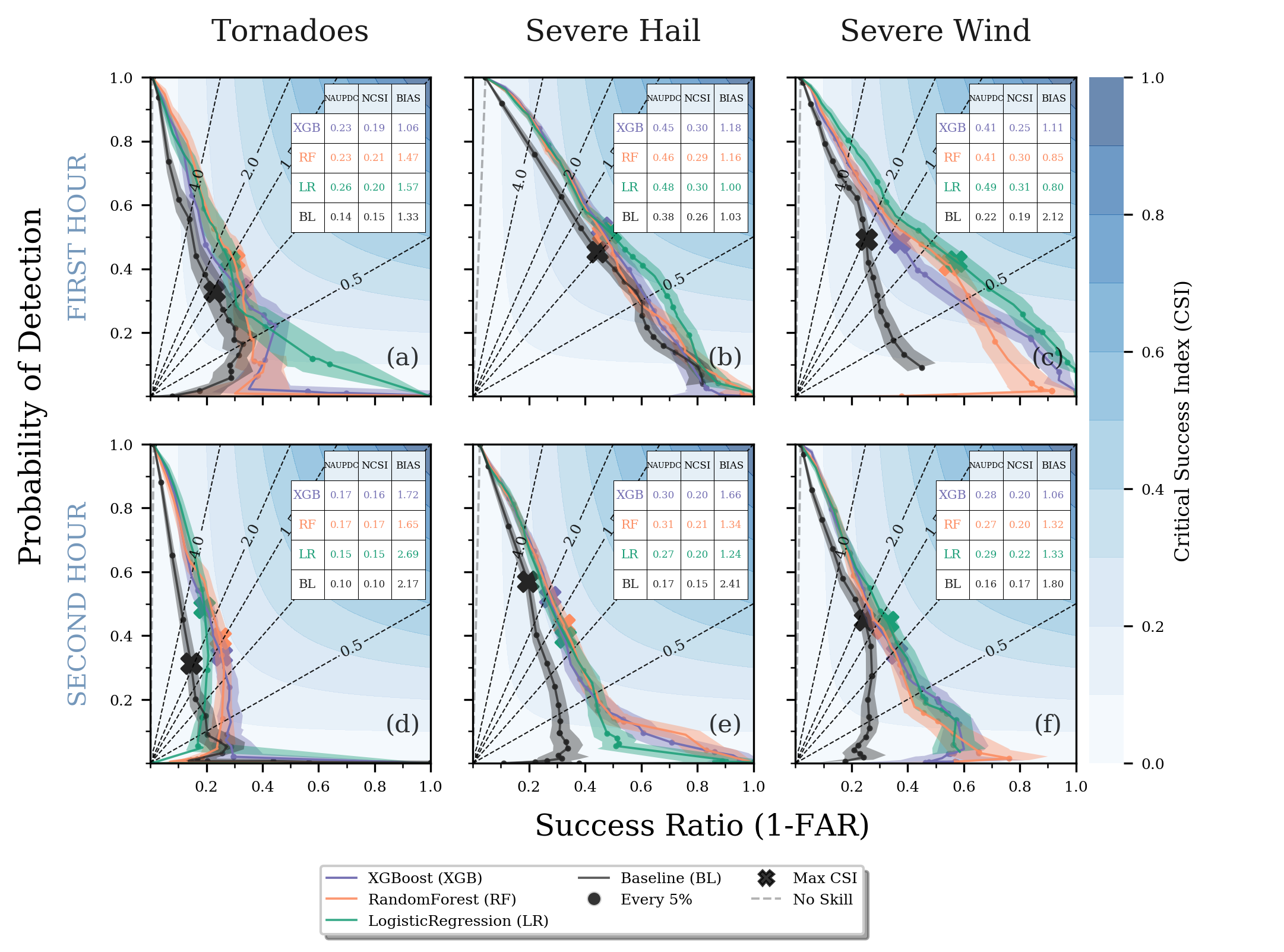}\\
  \caption{Same as in Fig.\ref{fig:roc_curves}, but for the performance diagram. The filled contours indicate the critical success index (CSI) while the dashed diagonal lines are the frequency bias. The dashed grey line indicates a no-skill classifier defined by equation \ref{eqn:sr_min}. The mean NAUPDC, NCSI, and frequency bias (BIAS) for each model are provided in the table in the upper right hand side of each panel. The maximum CSI is denoted on each curve with an X }\label{fig:perform_curves}
\end{figure*}

The performance diagrams are shown in Figure~\ref{fig:perform_curves}. For the FIRST HOUR dataset (e.g., examples with a lead time of 0-30, 5-35, ..., 60-90 min; Fig.\ref{fig:perform_curves}a,b,c), the three ML models produced higher NAUPDC and maximum NCSI for severe hail and wind (Fig.~\ref{fig:perform_curves}b,c) than for tornadoes (Fig.~\ref{fig:perform_curves}a).  This is unsurprising as the severe wind and hail events are more frequent than tornadoes, giving the ML more opportunities to learn from those examples. 
In addition, the processes governing hail growth and generation of strong near-surface winds are better resolved on a 3-km grid than the processes governing tornadogenesis, which is strongly influenced by small-scale processes in at least some cases \citet{Coffer+etal2017,Flournoy+etal2020}. For tornadoes and severe hail, the NAUPDC and maximum NCSI of the three ML models were fairly indistinguishable from one another (Fig.~\ref{fig:perform_curves}a,b), but for severe wind (Fig.~\ref{fig:perform_curves}c), the random forest and logistic regression models produced significantly higher maximum NCSI than XGBoost. Other than for the severe wind random forest and logistic regression model, the frequency bias associated with maximum NCSI is greater than 1 (Fig.~\ref{fig:perform_curves}a,b), which matches expectations for rare events \citep{Baldwin+Kain2006}. 

All three ML models significantly outperformed the UH baseline, but the magnitude of improvement varied with severe weather hazard. For tornadoes and especially severe wind, the ML predictions substantially improved upon the baseline. The superiority of the ML model severe wind forecasts is not surprising, as mid-level UH is less correlated with severe wind events (which are often produced by non-rotating storms) than with severe hail and tornado potential. The baseline predictions performed the best on severe hail, which is expected as mid-level UH is a proxy for supercells, which are the most prolific producer of severe hail \citep{Duda+Gallus2010} and especially significant severe hail \citet{Smith+etal2012}. This result aligns with Gagne et al. (~\citeyear{Gagne+etal2017}) who found that UH predictions of severe hail competed with the ML-based predictions.

The performance curves were degraded for the SECOND HOUR dataset (e.g., examples with a lead time of 65-95, 70-100, ..., 120-150 min; Fig.\ref{fig:perform_curves}d,e,f). The POD remained relatively unchanged for tornadoes, but the FAR increased, which decreased the NAUPDC and maximum NCSI. The increase in FAR also led to the maximum CSI occurring with an increased over-forecasting frequency bias (especially for logistic regression). The predictability of storm-scale features relevant to tornado prediction (e.g., mid- and low-level mesocyclones) is greatly diminished at later lead times \citep{Flora+etal2018} and therefore this degradation in skill is not surprising. For severe hail and wind, the changes in POD and FAR relative to FIRST HOUR compensated each other such that the maximum-CSI frequency bias remained slightly above one. The major exception is the XGBoost severe hail model, which suffered from over-forecasting bias in the FIRST HOUR dataset but in the SECOND HOUR dataset has a maximum-CSI frequency bias near 1 (1.08). The difference in performance between the UH baseline predictions and the three ML models is more pronounced in SECOND HOUR than FIRST HOUR suggesting that ML-based calibration of ensemble forecasts is more useful at later lead times. This result suggests that the ML models are learning enough useful information from the ensemble statistics at these later lead times to partly compensate the inevitable reduction in CAM forecast skill because of intrinsically limited storm-scale predictability.

\begin{figure*}[t]
  \noindent\includegraphics[width=38pc,angle=0]{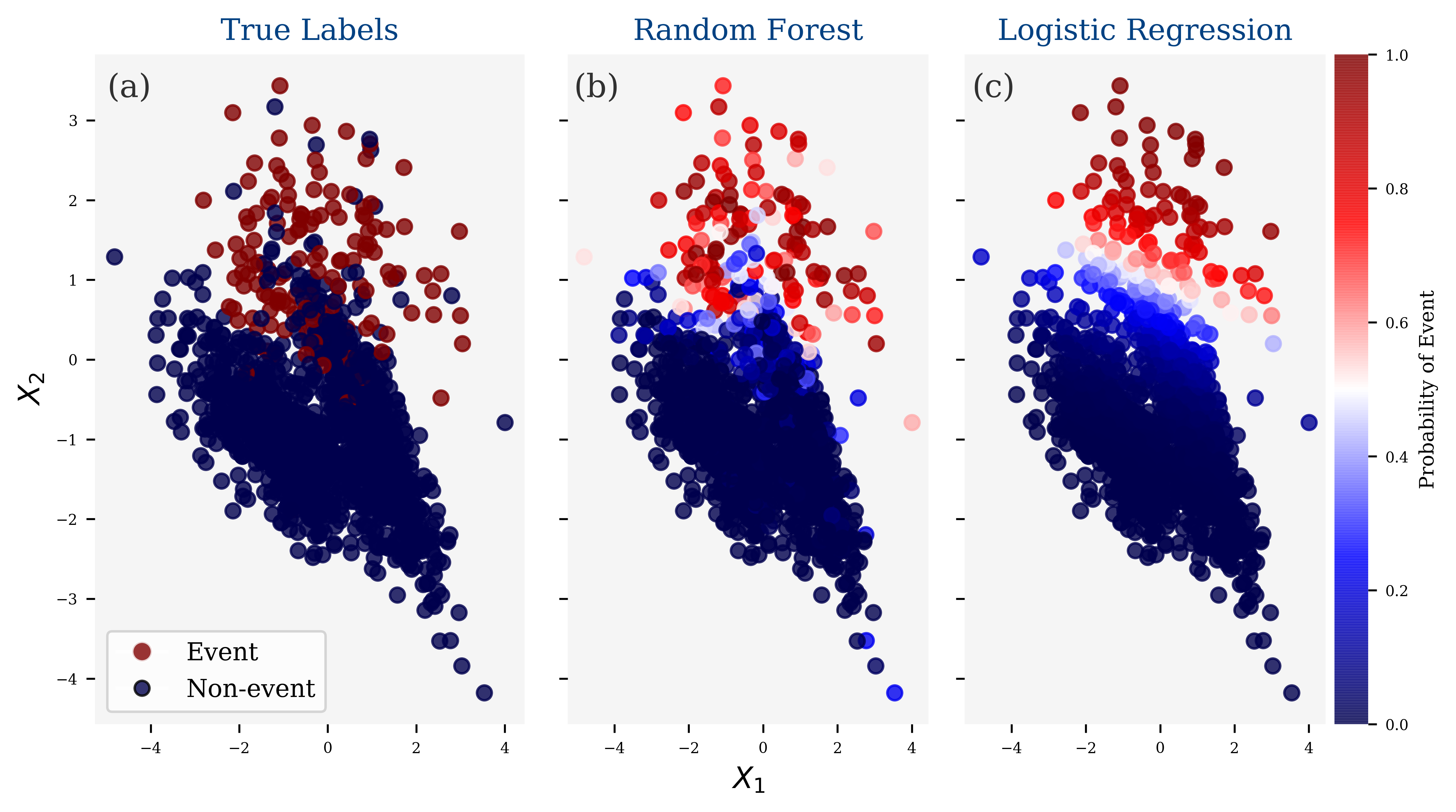}\\
  \caption{Illustration of predictions for a simple noisy 2D dataset (shown in a) from a random forest (shown in b) and logistic regression model (shown in c). }\label{fig:toydatasets}
\end{figure*}

For all three severe weather hazards, the logistic regression model has a significantly higher SR (lower FAR) at higher probability thresholds (lower right-hand portion of the diagram) than the other ML models, which explains the slightly higher mean NAUPDC values. To explain why logistic regression can produce fewer false alarms for higher confidence forecasts, Figure \ref{fig:toydatasets} illustrates how predictions from a random forest and logistic regression model compare for a simple noisy 2D dataset. 
A classic problem in ML is the trade-off between the bias and variance of a model. With a high-variance model, we risk over-fitting to noisy or unrepresentative training data. In contrast, a high-bias model is typically simpler and tends to underfit the training data, failing to capture important regularities. Tree-based methods partition the predictor space and produce predictions based on the local event frequency of the training dataset. If there is sufficient noise in the classification (e.g., ensemble storm tracks mislabeled as non-events because of missing storm reports), then the local event frequency could be unrepresentative of the true local event frequency. Though the tree-based method can produce skillful high confidence forecasts with noisier datasets (as seen in Fig. \ref{fig:toydatasets}b; \citealt{Hoekstra+etal2011}), they are high-variance models (more sensitive to random variations in the data) and can struggle near decision boundaries or in poorly sampled regions of the predictor space. For example, near point $(X_1;X_2) = (-1,1)$, the random forest probabilities do not reflect the uncertainty of the true labels and for points $X2>2$, the predictions have high confidence, but instances of unrepresentative uncertainty (e.g., the probability of point $(X_1;X_2) = (2, 2.5)$ is 50$\%$, but should be 100$\%$). Logistic regression is a lower-variance, higher-bias model compared to tree-based methods (since it is a linear model which may not sufficiently generalize a dataset) and so its predictions are not very sensitive to noisy labeling and rather, as we can see in Fig.~\ref{fig:toydatasets}, increase (or decrease) perpendicular to the linear decision boundary. Therefore, we propose that the logistic regression models in this study are producing fewer false alarms than tree-based models at higher probability thresholds since the tree-based methods are strongly impacted by the noisy labeling and are over-fitting the training dataset. However, the logistic regression models are not markedly better than the tree-based methods, so the tradeoff between bias and variance is still a relevant issue. It is likely that if the ensemble storm tracks were labeled better (improving the signal-to-noise ratio) then the tree-based methods would outperform logistic regression, since a linear decision boundary does not sufficiently generalize to the data.

\subsection{Attribute Diagrams}
The attribute diagram plots forecast probabilities against their conditional event frequencies \citep{wilks_2011}. Thus, the plot for a perfectly reliable forecast system will lie along the one-to-one line. Traditionally, the forecast probabilities are separated into equally spaced bins from which we compute the mean forecast probabilities and conditional event frequencies. The conditional event frequencies, however, can be sensitive to the bin interval, especially for smaller datasets. To address uncertainty in the conditional event frequencies, we computed the \say{consistency bars} from \citet{Brocker+Smith2007} which allows for an immediate interpretation of the confidence of the reliability of a prediction system. We can then assess reliability as the extent to which the conditional event frequencies fall within the consistency bars rather than strictly based on their distance from the diagonal. A common metric associated with the attribute diagram is the Brier skill score (BSS; \citealt{Hsu+Murphy1986}) where regions of positive and negative BSS can be delimited on the attribute diagram based on the climatological event frequency. The Brier skill score is defined as
\begin{equation}
    BSS = \frac{\Big[\frac{1}{K} \sum_{k=1}^{N} n_k (\overline{y_k} - \overline{y})^2\Big] - \Big[\frac{1}{N} \sum_{k=1}^{K} n_k (p_i - \overline{y_k})^2\Big]}{\overline{y}(1 - \overline{y})}
\end{equation}
where $p$ is the forecast probabilities, $y$ is the binary target variable, $K$ is the number of bins, $N$ is the number of examples, $n_k$ is the number of examples in the $k$th bin, and $\overline{y}$ is the climatological event frequency. The two terms in the numerator (from left to right) are known as resolution and reliability, respectively, while the denominator is the uncertainty term. Reliability measures how well the forecast probabilities correspond with the conditional event frequencies while resolution measures how the conditional event frequencies differ from the climatological event frequency. The uncertainty term refers to uncertainty in the observations and is independent of forecast quality. A positive BSS (resolution > reliability) means that the model is better than the baseline prediction (climatological event frequency). BSS is sensitive to class imbalance, but the authors are unaware of any methods that attempt to normalize BSS by the climatological event frequency.

\begin{figure*}[t]\
  \noindent\includegraphics[width=38pc,angle=0]{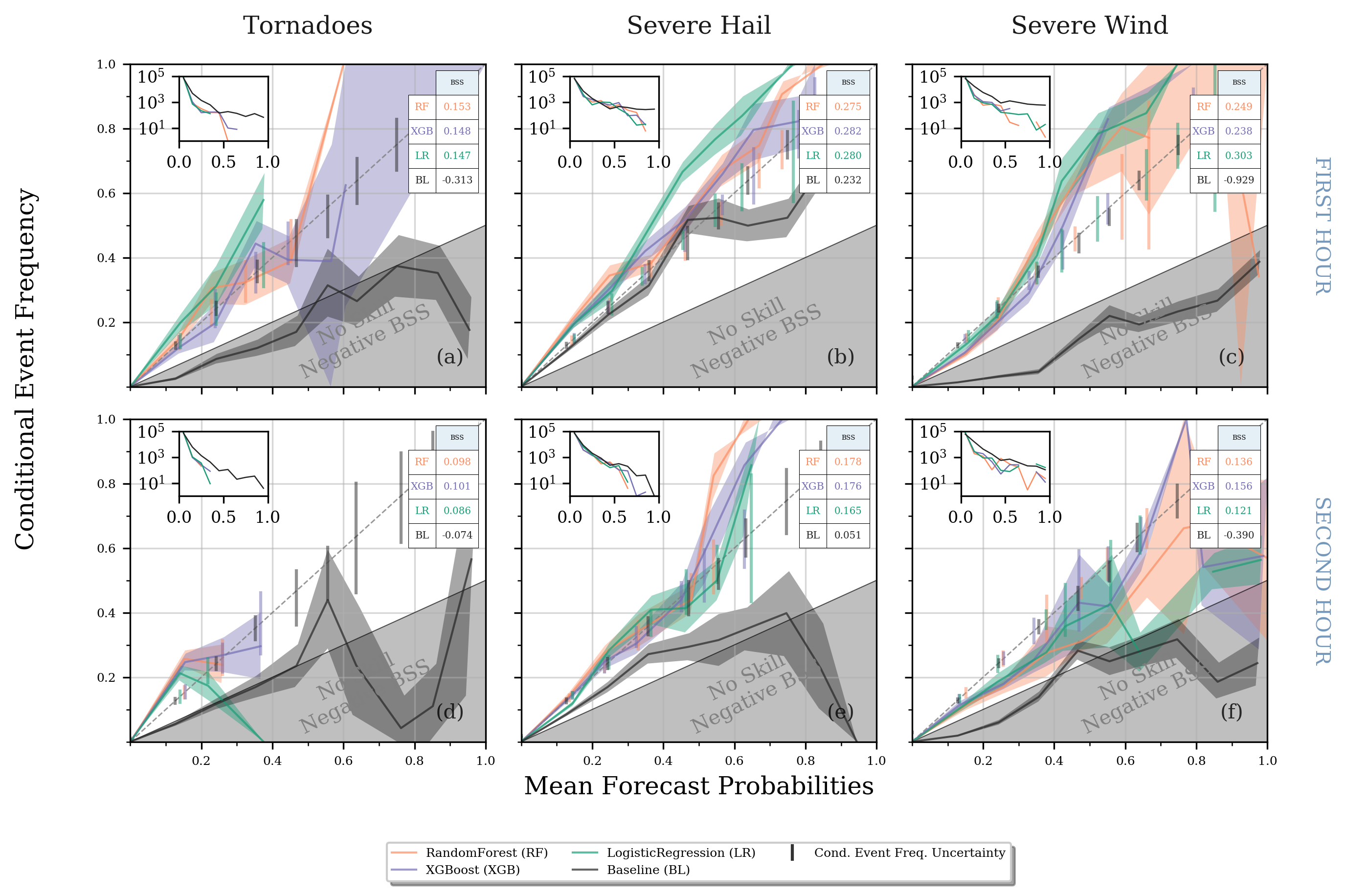}\\
  \caption{Same as in Fig.\ref{fig:roc_curves}, but for attribute diagrams. The bin increment of forecast probabilities is $10\%$. The inset figure is the forecast histogram for each model.  The dashed line represents perfect reliability while the grey region separates positive and negative Brier skill score (positive Brier skill score above the grey area). The vertical lines along the diagonal are the error bars for the observed frequency for each model in each bin based on the method in \citet{Brocker+Smith2007}. To limit figure crowding, error bars associated with an uncertainty of $>50\%$ for a given conditional observed frequency were omitted. The mean BSS for each model is provided in the table in the upper right hand side of each panel.}\label{fig:attr_curves}
\end{figure*}

The attribute diagram results are shown in Figure~\ref{fig:attr_curves}. For both lead time ranges, the severe hail and wind prediction were the most reliable (Fig.\ref{fig:attr_curves}b,c,e,f). 
The larger numbers of severe hail and wind events than tornado events in the training dataset likely contribute to increased reliability by improving the local event frequencies for the tree-based methods and the coefficients of the linear model in logistic regression. All three models produced reliable severe wind probabilities up to 40-50$\%$ with a small underforecasting bias for higher probabilities; no model produced forecast probabilities greater than 80$\%$ (Fig.\ref{fig:attr_curves}c). Severe hail probabilities for all three models were reliable up to 40$\%$ with a small over-forecasting bias for probabilities greater than 60$\%$ with probabilities up to 90$\%$ being produced. The under-forecasting bias was significantly higher for the logistic regression, which corresponds with the lower FAR at higher probabilities previously noted in the performance diagram (Fig.~\ref{fig:toydatasets}). Though the logistic regression model is less reliable than the tree-based models for severe wind and hail, its resolution is higher, which explains why its BSS is higher. The logistic regression model also produced the least reliable tornado predictions, exhibiting an under-forecasting bias, and only produced forecast probabilities up to 40$\%$. The tree-based models produced higher probabilities, but the uncertainty in the conditional event frequencies is too large to assess the forecast reliability at these higher probabilities.  The smaller forecast probabilities for tornadoes is not surprising for at least two reasons. First,  missing tornado reports \citep{Potvin+etal2019} coupled with the rarity of tornado events limits the ability of the ML models to learn subtle patterns in the data. Second, storm-scale predictability limits \citep{Flora+etal2018} prevents greater confidence in tornado likelihood, especially at later lead times. 

For all severe weather hazards, reliability and resolution were degraded for the SECOND HOUR dataset. The tornado probabilities are arguably reliable, but the maximum probability is between 30-40$\%$, which are fairly confident forecasts of such a rare event. For severe hail, the forecast probabilities remained relatively reliable, but the maximum forecast probability was significantly reduced, which lowered the BSS. The severe wind forecast probabilities for all three models became overconfident at later lead times (cf. Fig.~\ref{fig:attr_curves}c and Fig.~\ref{fig:attr_curves}f). 

For tornadoes and severe wind, the UH baseline was unreliable and unskillful at all lead times (underperformed climatology; Fig.\ref{fig:attr_curves}a,c,d,f). Reliability is possibly improved at a higher UH threshold, but then the ranking-based metrics would have suffered. This result highlights that the simple threshold method is likely over-fitting the training dataset and is suboptimal for capturing forecast uncertainty, which is similar to the result found in \citet{Sobash+etal2020}. The UH baseline was fairly reliable for severe hail, but the ML models were still significantly more reliable (Fig.\ref{fig:attr_curves}b).

%%%%%%%%%%%%%%%%%%%%%%%%%%%%%%%%%%%%%%%%%%%%%%%%
% CONCLUSIONS SECTION 
%%%%%%%%%%%%%%%%%%%%%%%%%%%%%%%%%%%%%%%%%%%%%%%%
\section{Conclusions}\label{conclusions}
The primary goal of Warn-on-Forecast is to provide human forecasters with short-term, storm-scale probabilistic severe weather guidance. Current CAM guidance can provide useful severe weather surrogates (e.g., updraft helicity), but it must be calibrated for individual severe weather hazards.  An emerging approach to solving this problem are ML models, which can easily incorporate many predictors, are well-suited for complex, noisy datasets, and have been shown to produce calibrated, skillful probabilistic guidance for a variety of meteorological phenomena. 

In this study, gradient-boosted classification trees, random forests, and logistic regression models were trained on WoFS forecasts from the 2017-2019 HWT-SFEs to predict which 30-min forecast storm tracks in the WoFS domain will produce a tornado, severe hail, and/or severe wind report up to lead times of 150 min. A novel ensemble storm track identification method inspired by \citet{Flora+etal2019} was used to extract ensemble statistics of intra-storm and environmental parameters. We labeled the ensemble storm tracks based on local storm reports, which, while error prone, are the best available severe weather database for individual hazards. We compared the ML predictions against the probability of mid-level UH exceeding a threshold that was tuned for each severe weather hazard. The primary conclusions are:
\begin{itemize}
    \item  The ML models produced significantly higher maximum Normalized Critical Success Indexs (NCSIs) and normalized area under the performance diagram than the UH baselines, especially at later lead times. This result is especially encouraging since observation-based severe weather prediction methods rapidly degrade beyond nowcasting lead times. 
     
    \item The ML models produced markedly more reliable predictions than the UH baselines, which were unreliable and produced negative BSS scores. 

    \item The ML models discriminated well (AUCs $>$ 0.9) for all three severe weather hazards up to a lead time of 150 min. 
    
    \item For a given severe weather hazard, the contingency table metrics for the three ML algorithms were fairly similar. The severe hail predictions had the highest NCSI while tornado predictions had the lowest NCSI especially at later lead times.
    
    \item Severe hail and wind predictions were more reliable than tornado predictions at all lead times. All three models produced fairly reliable hail and wind probabilities up to 50$\%$ while hail (wind) forecasts were under-confident (overconfident) for higher probabilities. At later lead times, severe hail forecast probabilities were reliable up to 60$\%$ while severe wind forecast probabilities became more overconfident.
    
\end{itemize}
While these results are promising, there are some limitations to this study that should be considered. First, since we are operating in an event-based framework, we are not correcting for instances when the WoFS fails to accurately analyze ongoing convection or exhibits biases in storm location. In future studies, we plan to adopt a hybrid gridpoint-based/event-based framework that, near missed storms, produces a complementary forecast that is largely based on environmental parameters. Second, the labelling of ensemble storm tracks was based on whether they contain a local storm report.  We showed that because of small spatial errors in forecast storm tracks, reports may fall just outside the boundary of an ensemble storm track. Given
these near-misses, and the spurious false alarms arising from missing storm reports, the verification results likely underestimate the potential ML skill. Third, we did not evaluate the ML models for different geographic regions (e.g., \citealt{Gagne+etal2014, Herman+Schumacher2018b, Sobash+etal2020}), diurnal times, or initialization time. The data in this study were largely sampled from the Great Plains (Fig.~\ref{fig:density_map}) so it is important to assess the ML model performance in other regions. In future work, we plan to expand upon the verification of the ML predictions to highlight any potential failure modes.

There are additional potential extensions of this work. First, though the ML predictions outperformed a competitive baseline, we did not compare against any preexisting method for predicting severe weather hazards (e.g., ProbSevere; \citealt{Cintineo+etal2014, Cintineo+etal2018}) nor did we explore using a more hazard-specific baseline like WRF-HAILCAST \citep{AdamsSelin+Ziegler2016,Adams-Selin+etal2019} for severe hail or model low-level wind gusts for severe wind.  To further assess the potential operational value of our prediction algorithms, and to increase forecaster trust in the algorithms, it will be necessary to evaluate the ML models  against existing methods. Second, the labels used in this study are based on error-prone local storm reports. It will be crucial as a community to address these deficiencies in severe weather reporting. An alternative to storm reports would be to use radar-observed azimuthal shear (\citealt{Smith+Elmore2004, Miller+etal2013, Smith+etal2016, Mahalik+etal2019}) as a proxy for severe weather, but this approach has its own limitations. Third, a robust verification of a complex, end-to-end automated ML system is nearly impossible as one cannot possibly account for a complete list of failure modes \citep{Doshi-Velez+Kim2017}. Therefore, human forecasters will continue to play a role in automated guidance (known as the human in the loop paradigm) and the combination of which has outperformed solely automated guidance for severe weather forecasting \citep{Karstens+etal2018}. Thus, to build human forecasters’ trust in ML predictions and maximize the use of automated guidance requires explaining the \say{why} of an ML model’s prediction in understandable terms and creating real-time visualizations of these methods \citep{Hoffman+etal2017, Karstens+etal2018}. In ongoing research, we are using several ML interpretation methods to examine whether the algorithms are learning physical relationships and developing real-time visuals that explain ML model predictions using methods such as Shapley Additive Explanations (SHAP; \citealt{Lundberg+Lee2016}). Fourth, the different ML algorithms were similarly skillful, but tended to over- and under-predict in different situations. The best forecast may therefore be a weighted average of the different ML predictions, just as ensembles outperform deterministic forecasts in numerical weather prediction. Ensemble approaches can also provide estimates of forecast uncertainty, which can improve the trustworthiness of ML methods.  Future work should therefore explore the use of ML model ensembles for severe weather prediction. Lastly, we did not evaluate the ability of the ML models to differentiate between severe weather hazards. In future work, it is worth exploring multi-class approaches (i.e., will a forecast storm produce hail or a tornado or both?).

In addition to the more traditional ML algorithms used in this study, we also plan to apply convolutional neural networks (CNNs; \citealt{LeCun+etal1990}) to WoFS forecasts to predict severe weather. The primary advantage of CNNs is that they can learn from spatial data and do not require manual predictor engineering. CNNs have also showed success for a variety of meteorological applications (e.g., Gagne et al.~\citeyear{Gagne+etal2019}; \citealt{Lagerquist+etal2019, Wimmers+etal2019, Lagerquist+etal2020}) and CNN interpretation techniques create metrics in the same space as the input spatial grids, making them easier to digest \citep{McGovern+etal2019_blackbox}. Given that CNN can encode spatial information, CNN techniques may also prove useful in the aforementioned hybrid gridpoint-based/event-based framework, especially in the situations where the WoFS does not contain a given observed storm. 

%%%%%%%%%%%%%%%%%%%%%%%%%%%%%%%%%%%%%%%%%%%%%%%%%%%%%
% ACKNOWLEDGMENTS
%%%%%%%%%%%%%%%%%%%%%%%%%%%%%%%%%%%%%%%%%%%%%%%%%%%%%
\acknowledgments
Funding was provided by NOAA/Office of Oceanic and Atmospheric Research under NOAA-University of Oklahoma Cooperative Agreement $\#$NA11OAR4320072, U.S. Department of Commerce. We thank Vanna Chmielewski for informally reviewing an early version of the manuscript and three anonymous reviewers for their comments, which substantially improved the manuscript. Valuable local computing assistance was provided by Gerry Creager, Jesse Butler, Jeff Horn, Karen Cooper, and Carrie Langston.

%%%%%%%%%%%%%%%%%%%%%%%%%%%%%%%%%%%%%%%%%%%%%%%%%%%%%
% DATA AVAILABILITY STATEMENT
%%%%%%%%%%%%%%%%%%%%%%%%%%%%%%%%%%%%%%%%%%%%%%%%%%%%%
\datastatement
The experimental WoFS ensemble forecast data used in this study is not currently available in a publicly accessible repository.  However, the data and code used to generate the results herein are available from the authors upon request.

\appendix
\appendixtitle{Derivation of Maximum Critical Success Index of a No-Skill System}
From \citet{Roebber2009}, the critical success index (CSI) can be defined as a function of success ratio ($s$) and probability of detection ($p$): 
\begin{equation}
    CSI = \frac{1}{s^{-1} + p^{-1} - 1} 
\end{equation}
Substituting the minimum success ratio for a no-skill system, into equation A1, we get 
\begin{equation}
    CSI = \frac{1}{\frac{1 - \pi + \pi p}{\pi p} + \frac{1}{p} - 1}.
\end{equation}
We then multiply the numerator and denominator by $\pi p$,
\begin{equation}
    CSI = \frac{\pi p}{1 - \pi + \pi p + \pi - \pi p}
\end{equation}
and then cancel the terms in the denominator to get the CSI of a no-skill system:
\begin{equation}
    CSI = \pi p.
\end{equation}
Based on equation A4, the maximum CSI of a no-skill system occurs for $p=1$ and is equal to climatological event frequency ($\pi$).

%%%%%%%%%%%%%%%%%%%%%%%%%%%%%%%%%%%%%%%%%%%%%%%%%%%%%%
% REFERENCES
%%%%%%%%%%%%%%%%%%%%%%%%%%%%%%%%%%%%%%%%%%%%%%%%%%%%%%%
\bibliographystyle{ametsoc2014}
\bibliography{references}
%%%%%%%%%%%%%%%%%%%%%%%%%%%%%%%%%%%%%%%%%%%%%%%%%%%%%%
% TABLES
%%%%%%%%%%%%%%%%%%%%%%%%%%%%%%%%%%%%%%%%%%%%%%%%%%%%%%%

\end{document}